# Preventive Audits for Data Applications Before Data Sharing in the Power IoT

Bohong Wang, *Graduate Student Member, IEEE*, Qinglai Guo, *Fellow, IEEE*, Yanxi Lin, and Yang Yu, *Member, IEEE*

*Abstract*—With the increase in data volume, more types of data are being used and shared, especially in the power Internet of Things (IoT). However, the processes of data sharing may lead to unexpected information leakage because of the ubiquitous relevance among the different data, thus it is necessary for data owners to conduct preventive audits for data applications before data sharing to avoid the risk of key information leakage. Considering that the same data may play completely different roles in different application scenarios, data owners should know the expected data applications of the data buyers in advance and provide modified data that are less relevant to the private information of the data owners and more relevant to the nonprivate information that the data buyers need. In this paper, data sharing in the power IoT is regarded as the background, and the mutual information of the data and their implicit information is selected as the data feature parameter to indicate the relevance between the data and their implicit information or the ability to infer the implicit information from the data. Therefore, preventive audits should be conducted based on changes in the data feature parameters before and after data sharing. The probability exchange adjustment method is proposed as the theoretical basis of preventive audits under simplified consumption, and the corresponding optimization models are constructed and extended to more practical scenarios with multivariate characteristics. Finally, case studies are used to validate the effectiveness of the proposed preventive audits.

## I. INTRODUCTION

AS the Internet of Things (IoT) has become more ubiquitous, a large number of information sensing devices, as well as large communication networks, interconnect people and objects in different times and spaces; thus, the volume of data has been sharply exploded, and the types of data have diversified. According to statistics and predictions from Statista, the volume of data created, captured, copied, and consumed worldwide may increase to more than 180 zettabytes by 2025, which is ten times larger than the value in 2015 [1]. Moreover, with the development of various technologies such as big data, cloud computing, and artificial intelligence, people's ability to process and analyze data has also greatly improved. People can collect and store massive amounts of data through high-capacity data storage and can mine and utilize them through high-performance data computing [2], thereby discovering the significant value contained in the data and actively transforming data assets into economic benefits through data sharing. Therefore, data have gradually become an important production factor that promotes economic growth and social development.

In the application scenarios of the power Internet of Things (IoT), data sharing is a necessary and effective way for entities to obtain more information, reduce uncertainty, and ultimately benefit from the data. For instance, in demand-side electricity transactions, electricity retailers serve as agents for end users to purchase electricity in the wholesale market, and sell electricity to end users in the retail market [3], [4]. However, electricity retailers usually do not fully know the real-time electric load demands of the end users. The lack of information about end users' demands generates risks to the decision-making of the electricity retailers when they purchase and sell electricity. To better mitigate the risks brought by the uncertainty, electricity retailers are willing to improve the forecasting of real-time electric load demand by obtaining more data about end users' electricity consumption behavior via data sharing. The data about end users' electricity consumption behavior are usually collected by data service providers, who engage market entities that convert raw data into data products or services [5], such as the Goods Center of Data Center, China Southern Power Grid [6], which analyzes electricity data to obtain enterprise credit indices, and provide data products to credit reporting agencies. They assist the credit reporting agencies in providing post-loan warnings, shell enterprise identification, credit anti-fraud and other services, and obtain the corresponding profits.

*A. Motivation*

However, considering that data, as an intangible asset, have significant differences compared to ordinary assets: in the absence of clear policy constraints, data sharing is similar to leasing ordinary assets; that is, data sharing grants data service providers the usufructuary right of data without transferring the ownership of data. Therefore, after obtaining data from data owners, data service providers can use the data, provide valuable information for other entities who want to achieve specific objectives through processing and analyzing the data, and then package the information into data products or services and benefit from the data products or services; however, they cannot continue reselling the data to other entities. For ordinary assets, the separation of the usufructuary right from ownership does

This work was supported by the National Key Research and Development Program of China 2020YFB2104500. *(Corresponding author: Qinglai Guo)*

B. Wang and Q. Guo are with the Department of Electrical Engineering, Tsinghua University, Beijing, China. E-mail: wbh19@mails.tsinghua.edu.cn; guoqinglai@tsinghua.edu.cn.

Y. Lin is with the Institute for Interdisciplinary Information Sciences, Tsinghua University, Beijing, China. E-mail: linyx22@mails.tsinghua.edu.cn.

Y. Yu is with the School of Economics and Management & Laboratory for Low-Carbon Intelligent Governance, Beihang University, Beijing, China. E-mail: yangyuenergy@gmail.com.

not have unexpected negative effects on the asset owners. For example, the transfer cost of the usufructuary right of a property can be fully determined in the rent through the agreement between the two entities in the property leasing process. In the process of using the property and realizing profits through property management, the impact of the tenant on the owner has already been included in the rent. However, the separation of the usufructuary right from the property ownership may have unexpected negative impacts on data owners. The value of usufructuary rights is difficult to determine due to the wide range of data application content involved. According to the smart meter data of residential users, and the user survey of the Commission for Electricity Regulator (CER) in Ireland, after data service providers have obtained electric load data from the smart meters of end users, they can infer some social characteristics about the end users, such as the end users' gender, employment, education, social class, internet access, etc. Through the processing and analysis of electric load data, data service providers can draw conclusions about the social characteristics of the end users, and are expected to further provide relevant information to other market entities, such as headhunters, estate agencies, and matchmaking agencies. In fact, in the process of data sharing, data owners only grant data service providers the usufructuary right of electric load data, but data service providers invisibly control the usufructuary right of the electric load data, and may leak some private characteristics of the end users, i.e., the data owners, and create unexpected negative effects to them. There are various ways to implement similar usufructuary rights. Before data sharing, it is difficult for data owners to strictly regulate on how the data service providers use their data or to conduct a traversal examination of possible ways in which the data service providers use data, and to reasonably determine the value of their usufructuary rights.

To prevent data service providers from using data inappropriately, which may harm the legitimate rights and interests of data owners in protecting their data privacy, it is necessary to conduct comprehensive preventive audits for data applications before sharing data. Preventive audits should analyze the correlation between the unshared data and the private characteristics of the data owners, and preprocess the unshared data to minimize the correlation. This will ensure that the data service providers obtain shared data which are processed through preventive audits, and therefore will find it difficult to infer the private characteristics of the data owners based on the shared data.

*B. Literature Review*

Data audits can commonly be found in the theoretical research field of information theory, while audits are rarely involved in the practical application of power systems and the IoT.

Chakraborty et al. [7] proposed a general application scenario where data owners delegate personal information to specific data receivers for processing. Data owners set whitelists and blacklists for personal information, and the ratio of mutual information to information entropy is used as a representation of privacy and utility indices. Asoodeh et al. [8] treated private information, nonprivate information, and display data as Markov chains, and constructed display data based on known private information and nonprivate information. The cases where private and nonprivate information are discrete random variables and continuous random variables are discussed separately. Rassouli et al. [9] assumed that data owners have access to both private information and useful data, and the correlation between useful data and private information is achieved through a leakage matrix. This paper fixes the mutual information of the disclosed data and private information at 0 and maximizes the mutual information of the disclosed data and useful data, and the maximum value is proven under the output perturbation model and the full information observation model. Zamani et al. [10] derived the function representation lemma proposed in [11] under the same scenario setting, which provided the condition for the mutual information of the disclosed data and private information to be 0 and the method for constructing disclosed data. This paper extended the assumption that the mutual information of the disclosed data and private information does not exceed a small positive number $\varepsilon$, derived the lower bound of the maximum mutual information of the disclosed data and useful data in general, and the expression of the maximum mutual information in specific settings. Based on the strong function representation lemma proposed in [12], this paper also obtained privacy-utility trade-off optimization results after expanding the assumptions. Zamani et al. [13] extended the assumption to measure the privacy-utility trade-off for all possible values of the disclosed data, thereby obtaining the asymptotic optimal bound for the privacy-utility trade-off corresponding to the disclosed data when the private information is a deterministic function of the useful data. Zamani et al. [14] further extended the assumption that the leakage matrix is an irreversible matrix, and obtained a method for constructing the disclosed data, as well as the maximum mutual information of the disclosed data and useful data in specific settings. Zamani et al. [15] further extended the composition of the private information and useful data, stating that multiple pieces of private information and multiple pieces of useful data belong to multiple data owners, and that the mutual information of multiple pieces of the private information and disclosed data is not greater than a small positive number $\varepsilon$; then, they constructed disclosed data to maximize the mutual information of the multiple pieces of the useful data and disclosed data.

*C. Main Contributions*

There are significant differences between the existing research and the main content in this paper. Compared with the commonly used audit in Fig. 1, a preventive audit for data applications are constructed in Fig. 2. Data owners have private characteristic $X^{(1)}$, nonprivate characteristic $X^{(2)}$, and unshared data $Y$. The correlation between unshared data $Y$ and private characteristic $X^{(1)}$ and between nonprivate characteristic $X^{(2)}$ are achieved through leakage matrices, and private characteristic $X^{(1)}$ and nonprivate characteristic $X^{(2)}$ can be inferred to a certain extent from unshared data $Y$.

Unlike existing research, this paper fully considers the complexity of unshared data $Y$ and the universal applicability of preventive audits for data applications. Unshared data $Y$ may be associated with multiple private and nonprivate characteristics of the data owners, where the private characteristics refer to the information that the data owners do not wish to be inferred or used for other purposes, such as personal details, health status, marital status, etc.; nonprivate characteristics refer to the information that the data owners agree to, or are not concerned about, its application, such as the number of household appliances and cooking types. Directly providing the unshared data $Y$ to the data service providers may result in the accidental disclosure of some private information of the data owners; providing fully desensitized data $Y$ to data service providers for disclosure may affect the legitimate use of the shared data. Therefore, it is necessary to establish reasonable preventive audits for data applications, ensuring that the private characteristics are protected and that the nonprivate characteristics are not affected. For data sharing between the data service providers and data owners, a preventive audit is designed to avoid the data service providers inferring and inappropriately using private characteristics, while ensuring that the data service providers use nonprivate characteristics normally. The key lies in using mutual information as the correlation index between the unshared data, private characteristics, and nonprivate characteristics, and adjusting the unshared data to achieve a privacy-utility trade-off.

The main contributions of this paper are listed as follows:
a) From the perspective of semantic information theory, mutual information is set as the data feature parameter to depict the correlation between data and characteristics in the privacy-utility trade-off. In addition, at different stages of development in the data sharing frameworks, different types of prevent audits are discussed.
b) Probability exchange adjustment methods with constant and variable probability distributions are proposed. To meet the privacy requirements of data owners and the utility requirements of data service providers, the mutual information pair is inversely adjusted. Based on information theory, three propositions for the upper and lower bounds of the mutual information pair are proven.
c) An optimization model and its extended version of preventive audits for data applications are proposed. Based on the probability exchange adjustment method, the optimization model is constructed with the weighted sum of mutual information as the objective function. Besides, the private and nonprivate characteristics are extended to multivariate random variables to be more practical.

The remainder of this paper is organized as follows. In Section II, information-related concepts about preventive audits for data applications are introduced, and privacy-utility trade-offs against the background of power IoT are constructed. In Section III, probability exchange adjustment methods with constant and variable probability distributions are described and derived in detail. In Section IV, probability exchange adjustment methods are embedded into optimization models and extended to a general form that is feasible for multivariate private and nonprivate characteristics. Case studies with actual electric load data and the end users' social characteristics are implemented in Section V. Finally, Section VI concludes this paper and introduces potential research interests for future work.

## II. INFORMATION-RELATED CONCEPTS ON PREVENTIVE AUDIT FOR DATA APPLICATIONS

### A. Data Feature Parameters

In the existing research, Shannon's information entropy or differential entropy is often used as the data feature parameter to depict the uncertainty of data [16], [17]. However, according to Shannon's information theory, the uncertainty depicted by information entropy is limited to the syntactic level, that is, the communication errors caused by source symbols and channel noise during communication transmission are considered [18], thus it is inappropriate to directly use information entropy or differential entropy to depict the uncertainty on the semantic level. Fortunately, with the development of information theory, Shannon's information theory has been extended to the semantic level and pragmatic levels [19], [20].

The semantic background of data has received particular attention in the field of communication research. Due to the problems of channel capacity approaching Shannon's limit and source coding efficiency approaching Shannon's information entropy in existing 5G communication systems [21], semantic communication will become one of the key technologies in future 6G communication systems. Compared to Shannon's information entropy, which measures the syntactic uncertainty of data, the semantic entropy, which measures the semantic uncertainty of data, has more applications in the field of communication. In the fields of power systems and IoT, when using data to assist entities in decision-making, it is also necessary to improve the efficiency of data sharing and reduce the corresponding communication costs through data semantic interactions. According to existing research on semantic entropy [22], [23], most of the expressions of semantic uncertainty still refer to Shannon's information entropy [24], [25] and are defined as semantic entropy by replacing the probability of information entropy with the probability of specific tasks, based on a specific knowledge basis, which indicates that the expression of information entropy is still applicable in semantic communication. Combining the expression of information entropy with the probability distribution of semantics is a reasonable expression of semantic entropy to illustrate the semantic uncertainty of data.

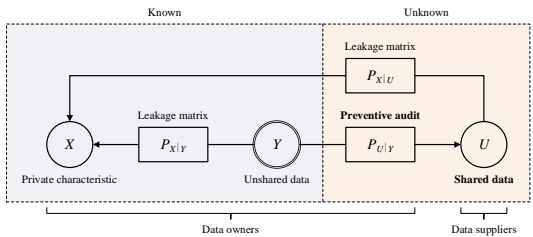

Fig. 1. Audit framework for data applications used in existing literature.

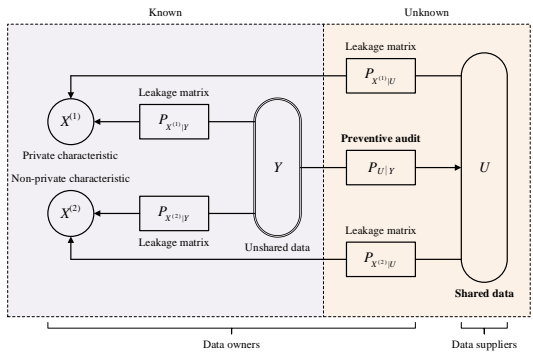

Fig. 2. Preventive audit framework of data applications proposed in this paper.

In this paper, for data sharing in power IoT data, as shown in Fig. 3, there is semantic uncertainty derived from data in specific organizational methods and explanatory models, such as multiple types of electric load statistical data derived from the same set of electric load data. Therefore, the expression of semantic entropy adopts an expression similar to the general semantic entropy proposed in [26]. Furthermore, semantic mutual information, which is abbreviated as "mutual information", is also derived from the expression of semantic entropy.

### B. Random Variables for Preventive Audits

For electricity retailers, more information about end users can help them better understand end users' electricity consumption behavior and the underlying reasons, thereby forecasting their end users' electric load more accurately. Electric load data collected from smart meters are commonly used in load forecasting, while those data may be related to different social characteristics of the end users, such as the end users' gender, employment, education, social class, internet access, family members, and house type. According to the description of private and nonprivate characteristics in Section I, data owners can classify social characteristics as private or nonprivate characteristics based on their own cognition.

Considering practical applications, end users' social characteristics are mostly represented by discrete variables rather than continuous variables. For example, an end user's gender usually has only two options, male and female, and an end user's education situation usually has only several countable options, such as primary school, junior high school, senior high school, undergraduate, or graduate. Even characteristics with numerical values, such as family members and housing usage duration, can be divided into several countable options or intervals, and thus present as discrete variables. Therefore, both end users' private and nonprivate characteristics are considered as discrete variables in this paper. In information theory, the information entropy of discrete random variables and the differential entropy of continuous random variables are not different expressions of the same concept but are different concepts with similar expressions [27]; thus, they are not suitable for simultaneous use in a single scenario. In addition, the calculations of concepts such as cross entropy, conditional entropy, and mutual information between discrete and continuous random variables are relatively complex and may deviate from their physical meanings [28]. Therefore, unshared data and shared data in the power IoT, represented by electricity load data that data service providers hope to obtain from data owners, are also considered discrete random variables, and their values are divided into several countable intervals.

$X^{(1)}$, $X^{(2)}$, $Y$, and $U$ are used to represent private characteristic, nonprivate characteristic, unshared data, and shared data, respectively. They are discrete random variables and are suitable for using information entropy to depict their uncertainty levels.

### C. Theoretical Value Range of the Mutual Information Pair

To achieve the objective of a preventive audit, that is, to avoid data service providers inferring private characteristic $X^{(1)}$ through unshared data $Y$, and to not prevent them from inferring nonprivate characteristic $X^{(2)}$ through unshared data $Y$, mutual information is the most noteworthy index. According to the physical meaning of the mutual information, the smaller the mutual information of unshared data $Y$ and private characteristic $X^{(1)}$, the less information that overlaps between unshared data $Y$ and private characteristic $X^{(1)}$, and the more ambiguous the information about private characteristic $X^{(1)}$ can be inferred from unshared data $Y$; the greater the mutual information of unshared data $Y$ and nonprivate characteristic $X^{(2)}$, the more information that overlaps between unshared data $Y$ and nonprivate characteristic $X^{(2)}$, and the more significant the information about nonprivate characteristic $X^{(2)}$ can be inferred from unshared data $Y$. Therefore, from the perspective of information theory, the requirement of the preventive audit can be transformed to decrease the mutual information of unshared data $Y$ and private characteristic $X^{(1)}$, and increase the mutual information of unshared data $Y$ and nonprivate characteristic $X^{(2)}$.

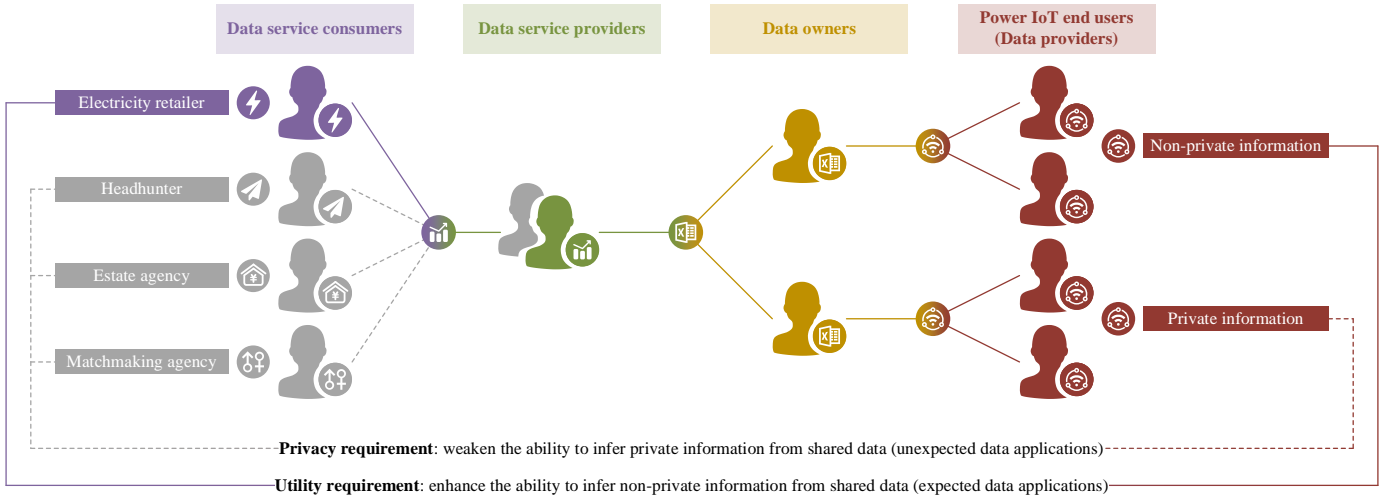

Fig. 3. Application framework of the preventive audit for data applications and the privacy and utility requirements.

Based on the expression of mutual information, there are theoretical upper and lower bounds on the two indices of mutual information as shown below:

$$\begin{cases} 0 \leq I(X^{(1)};Y) \leq H(X^{(1)}) \\ 0 \leq I(X^{(2)};Y) \leq H(X^{(2)}) \end{cases} \quad (1)$$

Due to the coupling of the two indices of mutual information, their values are not independent; the difference between the two indices of mutual information can be expressed as

$$I(X^{(2)};Y) - I(X^{(1)};Y) = I(X^{(2)};Y \mid X^{(1)}) - I(X^{(1)};Y \mid X^{(2)}) \quad (2)$$

Similarly, there are also theoretical upper and lower bounds on the two indices of conditional mutual information in Eq. (2), which are not related to the unshared data $Y$, as follows:

$$\begin{cases} 0 \leq I(X^{(2)};Y \mid X^{(1)}) \leq H(X^{(2)} \mid X^{(1)}) \\ 0 \leq I(X^{(1)};Y \mid X^{(2)}) \leq H(X^{(1)} \mid X^{(2)}) \end{cases} \quad (3)$$

Therefore, the difference between the two indices of mutual information can be expressed as

$$-H(X^{(1)} \mid X^{(2)}) \leq I(X^{(2)};Y) - I(X^{(1)};Y) \leq H(X^{(2)} \mid X^{(1)}) \quad (4)$$

Using the mutual information of unshared data $Y$ and private characteristic $X^{(1)}$ as the horizontal axis, and the mutual information of unshared data $Y$ and nonprivate characteristic $X^{(2)}$ as the vertical axis, the theoretical value range of the mutual information pair can be depicted as the gray area in Fig. 4.

*D. Privacy-Utility Trade-off*

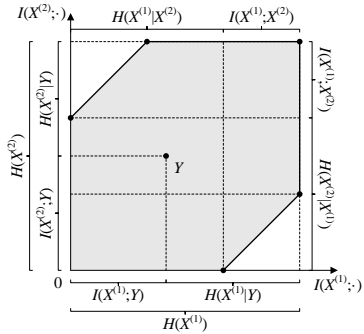

Fig. 4. Theoretical value range of the mutual information pair.

The preventive audit simultaneously focuses on the two indices of mutual information to facilitate the analysis of the process of converting unshared data $Y$ into shared data $U$. The privacy-utility trade-off is used to treat the mutual information of unshared data $Y$ or shared data $U$ and private characteristic $X^{(1)}$ as the privacy index, and the mutual information of unshared data $Y$ or shared data $U$ and nonprivate characteristic $X^{(2)}$ as the utility index. A preventive audit requires negative changes in the privacy index and nonnegative changes in the utility index, which corresponds to the light red and dark red areas in Fig. 5.

Considering that the privacy requirements of the data owners are often based on their concerns about the leakage of private characteristic $X^{(1)}$, rather than simply improving the privacy protection level relative to unshared data $Y$, the utility requirements of the data service providers are often based on their expectations of using unshared data $Y$ to fully complete load forecasting rather than just avoiding the negative impact of utility levels relative to unshared data $Y$. Therefore, the privacy and utility requirements should be more stringent. Furthermore, $\varepsilon$ and $\eta$ can be set as the margins of the privacy and utility requirements, respectively, where $\varepsilon$ and $\eta$ are both positive numbers, corresponding to the dark red area in Fig. 5.

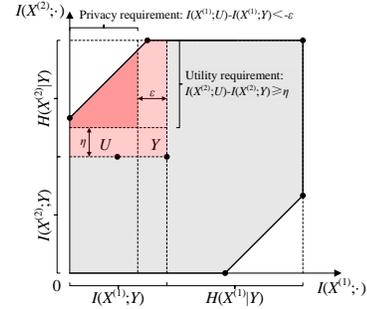

Fig. 5. Privacy-utility trade-off between the privacy requirement of data owners and utility requirement of data service providers.

In the privacy-utility trade-off, the implementation of preventive audits varies depending on the privacy requirements of the data owners and the utility requirements of the data service providers. At different stages of development in the data sharing frameworks, the preventive audits can be roughly divided into three types:

a) In the early stage, the data products or services are often simple in form and small in scale, and the data owners need to meet the requirements of the data service providers as much as possible. Therefore, data sharing is a buyer's market dominated by the data service providers, and the preventive audits mainly consider their utility requirements with constraints on the privacy requirements of the data owners;

b) In the middle stage, large centralized and specialized data owners, e.g., data platforms, grasp the core channels for obtaining raw data and develop a large number of data products or services. Therefore, data sharing is a seller's market dominated by the data owners, and the preventive audits mainly consider the privacy requirements of the data owners with constraints on the utility requirements of the data service providers;

c) In the later stage, both the data service providers and data owners become more mature, and the

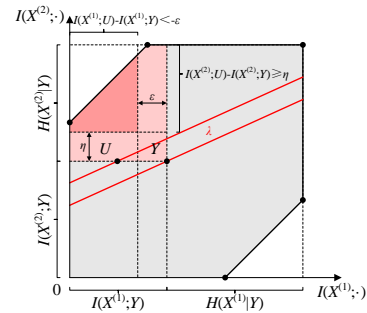

Fig. 6. Privacy-utility trade-off in the later stage of development in the data sharing framework of power IoT.

homogenization degree of the data products and services is relatively high. Therefore, data sharing gradually moves towards a perfectly competitive market, and the preventive audits may be executed by third-party regulatory agencies that simultaneously consider both the utility requirements of the data service providers and the privacy requirements of the data owners by setting a certain weight $\lambda$, as shown in Fig. 6.

## III. PROBABILITY EXCHANGE ADJUSTMENT METHOD

To clearly demonstrate the preventive audit process, a theoretical analysis is first conducted based on simplified assumptions. It is assumed that the private characteristic $X^{(1)}$ and nonprivate characteristic $X^{(2)}$ involved in the preventive audit are both binary discrete random variables, that is, discrete random variables with only two possible values. For example, if the private characteristic $X^{(1)}$ represents "family members", the original values may include 1, 2, 3, 4, etc. These values can be simplified into two groups of "≤2" and ">2", thereby simplifying the meaning of the characteristic while remaining unchanged. Unshared data $Y$ are represented by electric load data, and the possible values are divided into intervals. It is assumed that there are $N$ possible intervals for unshared data $Y$, namely, $Y_i$, $i \in \mathcal{Y}=\{1, 2, …, N\}$. Therefore, there are $4N$ possible values for the joint distribution of unshared data $Y$, private characteristic $X^{(1)}$, and nonprivate characteristic $X^{(2)}$. By setting the 4 value combinations of private characteristic $X^{(1)}$ and nonprivate characteristic $X^{(2)}$ with subscripts $a$, $b$, $c$, and $d$, the probability of the joint distribution can be divided into four groups of $N$ expressions, as shown below:

$$\begin{cases} \sum_{i=1}^{N} P_{Yi,a} = \sum_{i=1}^{N} P\left(Y=Y_i, X^{(1)}=X_1^{(1)}, X^{(2)}=X_1^{(2)}\right) \\ \sum_{i=1}^{N} P_{Yi,b} = \sum_{i=1}^{N} P\left(Y=Y_i, X^{(1)}=X_1^{(1)}, X^{(2)}=X_2^{(2)}\right) \\ \sum_{i=1}^{N} P_{Yi,c} = \sum_{i=1}^{N} P\left(Y=Y_i, X^{(1)}=X_2^{(1)}, X^{(2)}=X_1^{(2)}\right) \\ \sum_{i=1}^{N} P_{Yi,d} = \sum_{i=1}^{N} P\left(Y=Y_i, X^{(1)}=X_2^{(1)}, X^{(2)}=X_2^{(2)}\right) \end{cases} \quad (5)$$

Correspondingly, the information entropy of unshared data $Y$, the conditional information entropy of unshared data $Y$ conditioned on private characteristic $X^{(1)}$, and the conditional information entropy of unshared data $Y$ conditioned on nonprivate characteristic $X^{(2)}$ are as follows:

$$H(Y) = -\sum_{i=1}^{N} P_{Yi} \log P_{Yi} = -\sum_{i=1}^{N} \left(\sum_{r \in \mathcal{P}} P_{Yi,r}\right) \log \sum_{r \in \mathcal{P}} P_{Yi,r} \quad (6)$$

$$H(Y \mid X^{(1)}) = -\sum_{i=1}^{N} \left(\sum_{r=a,b} P_{Yi,r}\right) \log \frac{\sum_{r=a,b} P_{Yi,r}}{p_1} \\ -\sum_{i=1}^{N} \left(\sum_{r=c,d} P_{Yi,r}\right) \log \frac{\sum_{r=c,d} P_{Yi,r}}{1-p_1} \quad (7)$$

$$H(Y \mid X^{(2)}) = -\sum_{i=1}^{N} \left(\sum_{r=a,c} P_{Yi,r}\right) \log \frac{\sum_{r=a,c} P_{Yi,r}}{p_2} \\ -\sum_{i=1}^{N} \left(\sum_{r=b,d} P_{Yi,r}\right) \log \frac{\sum_{r=b,d} P_{Yi,r}}{1-p_2} \quad (8)$$

where

$$p_1 = P\left(X^{(1)} = X_1^{(1)}\right) \quad (9)$$

$$p_2 = P\left(X^{(1)} = X_1^{(2)}\right) \quad (10)$$

$\mathcal{P}=\{a, b, c, d\}$ is the set of different value combinations of private characteristic $X^{(1)}$ and nonprivate characteristic $X^{(2)}$.

### A. Probability Exchange Adjustment Method with Constant Probability Distributions

According to the expression of mutual information, to decrease the mutual information of unshared data $Y$ and private characteristic $X^{(1)}$, without decreasing the mutual information of unshared data $Y$ and nonprivate characteristic $X^{(2)}$, the conditional information entropy of unshared data $Y$ conditioned on private characteristic $X^{(1)}$ can increase, and the conditional information entropy of unshared data $Y$ conditioned on nonprivate characteristic $X^{(2)}$ cannot increase, while keeping the information entropy of unshared data $Y$ unchanged. Therefore, during the process of adjusting unshared data $Y$ to shared data $U$, the preventive audit can achieve the above objectives by adjusting the probability distribution of $Y_{i,r}$ while ensuring that the probability distribution of $Y_i$ remains unchanged, as shown below:

$$P_{Uj,r} - P_{Yj,r} = P_{Yj,s} - P_{Uj,s} = P_{Uk,s} - P_{Yk,s} = P_{Yk,r} - P_{Uk,r} = \delta \quad (11)$$

where $j, k \in \mathcal{Y}$, $r, s \in \mathcal{P}$, $\delta$ is the probability adjustment value.

Based on the comparison of the different information theory indices regarding unshared data $Y$ and shared data $U$, it can be found that

a) When $r$ and $s$ are $a$ and $b$ or $c$ and $d$, the conditional information entropy of unshared data $Y$ conditioned on private characteristic $X^{(1)}$ remains unchanged, while the conditional information entropy of unshared data $Y$ conditioned on nonprivate characteristic $X^{(2)}$ changes, as shown in Fig. 7:

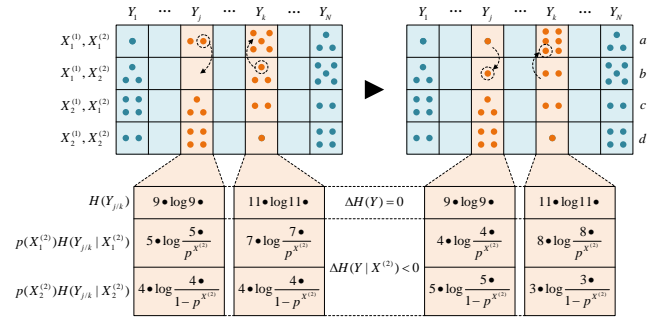

Fig. 7. Increased mutual information of unshared data $Y$ and nonprivate characteristic $X^{(2)}$ after the probability exchange adjustment method with constant probability distribution.

b) When $r$ and $s$ are $a$ and $c$ or $b$ and $d$, the conditional information entropy of unshared data $Y$ conditioned on private characteristic $X^{(1)}$ changes, while the conditional information entropy of unshared data $Y$ conditioned on nonprivate characteristic $X^{(2)}$ remains unchanged, as shown in Fig. 8:

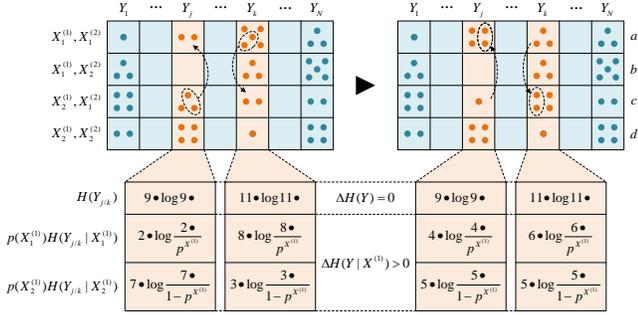

Fig. 8. Decreased mutual information of unshared data $Y$ and private characteristic $X^{(1)}$ after the probability exchange adjustment method with constant probability distribution.

c) When $r$ and $s$ are $a$ and $d$ or $b$ and $c$, the conditional entropy of unshared data $Y$ conditioned on private characteristic $X^{(1)}$ and the conditional information entropy of unshared data $Y$ conditioned on nonprivate characteristic $X^{(2)}$ both change, as shown in Fig. 9:

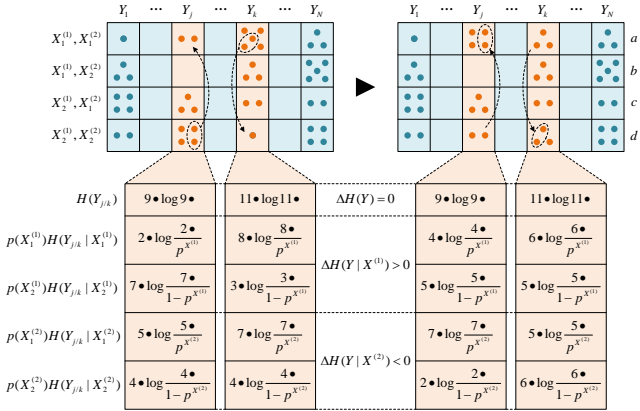

Fig. 9. Inversely changed mutual information pair after the probability exchange adjustment method with constant probability distribution.

Therefore, through the probability exchange adjustment method, the mutual information changes expected by the preventive audit can be achieved. Due to the equivalence between the information entropy of unshared data $Y$ and the information entropy of shared data $U$, the increase in the conditional information entropy of unshared data $Y$ conditioned on private characteristic $X^{(1)}$ equals the decrease in the mutual information of private characteristic $X^{(1)}$ and unshared data $Y$; the decrease in the conditional information entropy of unshared data $Y$ conditioned on nonprivate characteristic $X^{(2)}$ equals the increase in the mutual information of nonprivate characteristic $X^{(2)}$ and unshared data $Y$.

The conditional information entropy change of unshared data $Y$ conditioned on private characteristic $X^{(1)}$ and the conditional information entropy change of unshared data $Y$ conditioned on nonprivate characteristic $X^{(2)}$ are:

$$\Delta H_{jk}^{X^{(1)}}(\delta) = \Delta H_{ac,jk}^{X^{(1)}} = \Delta H_{ad,jk}^{X^{(1)}} = \Delta H_{bc,jk}^{X^{(1)}} = \Delta H_{bd,jk}^{X^{(1)}}$$

$$= \sum_{r=a,b} P_{Yj,r} \log \frac{\sum_{r=a,b} P_{Yj,r}}{\sum_{r=a,b} P_{Yj,r} + \delta} + \sum_{r=c,d} P_{Yj,r} \log \frac{\sum_{r=c,d} P_{Yj,r}}{\sum_{r=c,d} P_{Yj,r} - \delta}$$

$$+ \sum_{r=a,b} P_{Yk,r} \log \frac{\sum_{r=a,b} P_{Yk,r}}{\sum_{r=a,b} P_{Yk,r} - \delta} + \sum_{r=c,d} P_{Yk,r} \log \frac{\sum_{r=c,d} P_{Yk,r}}{\sum_{r=c,d} P_{Yk,r} + \delta} \quad (12)$$

$$+ \delta \log \frac{\sum_{r=c,d} P_{Yj,r} - \delta}{\sum_{r=a,b} P_{Yj,r} + \delta} \cdot \frac{\sum_{r=a,b} P_{Yk,r} - \delta}{\sum_{r=c,d} P_{Yk,r} + \delta}$$

$$\Delta H_{jk}^{X^{(2)}}(\delta) = \Delta H_{ab,jk}^{X^{(2)}} = \Delta H_{ad,jk}^{X^{(2)}} = \Delta H_{cb,jk}^{X^{(2)}} = \Delta H_{cd,jk}^{X^{(2)}}$$

$$= \sum_{r=a,c} P_{Yj,r} \log \frac{\sum_{r=a,c} P_{Yj,r}}{\sum_{r=a,c} P_{Yj,r} + \delta} + \sum_{r=b,d} P_{Yj,r} \log \frac{\sum_{r=b,d} P_{Yj,r}}{\sum_{r=b,d} P_{Yj,r} - \delta}$$

$$+ \sum_{r=a,c} P_{Yk,r} \log \frac{\sum_{r=a,c} P_{Yk,r}}{\sum_{r=a,c} P_{Yk,r} - \delta} + \sum_{r=b,d} P_{Yk,r} \log \frac{\sum_{r=b,d} P_{Yk,r}}{\sum_{r=b,d} P_{Yk,r} + \delta} \quad (13)$$

$$+ \delta \log \frac{\sum_{r=b,d} P_{Yj,r} - \delta}{\sum_{r=a,c} P_{Yj,r} + \delta} \cdot \frac{\sum_{r=a,c} P_{Yk,r} - \delta}{\sum_{r=b,d} P_{Yk,r} + \delta}$$

where the domains of $\delta$ are respectively

$$-\min\left(\sum_{r=a,b} P_{Yj,r}, \sum_{r=c,d} P_{Yk,r}\right) \leq \delta \leq \min\left(\sum_{r=a,b} P_{Yk,r}, \sum_{r=c,d} P_{Yj,r}\right) \quad (14)$$

$$-\min\left(\sum_{r=a,c} P_{Yj,r}, \sum_{r=b,d} P_{Yk,r}\right) \leq \delta \leq \min\left(\sum_{r=a,c} P_{Yk,r}, \sum_{r=b,d} P_{Yj,r}\right) \quad (15)$$

The values of $\Delta H_{jk}^{X^{(1)}}$ and $\Delta H_{jk}^{X^{(2)}}$ at $\delta=0$ are both 0. Considering that $0\log 0$ is defined as 0 in the information theory calculation, the conditional information entropy will not change before the preventive audit:

Then, the derivatives of $\Delta H_{jk}^{X^{(1)}}$ and $\Delta H_{jk}^{X^{(2)}}$ with respect to $\delta$ are as follows:

$$\frac{d\Delta H_{jk}^{X^{(1)}}}{d\delta} = \log \frac{\sum_{r=c,d} P_{Yj,r} - \delta}{\sum_{r=a,b} P_{Yj,r} + \delta} \cdot \frac{\sum_{r=a,b} P_{Yk,r} - \delta}{\sum_{r=c,d} P_{Yk,r} + \delta} \quad (16)$$

$$\frac{d\Delta H_{jk}^{X^{(2)}}}{d\delta} = \log \frac{\sum_{r=b,d} P_{Yj,r} - \delta}{\sum_{r=a,c} P_{Yj,r} + \delta} \cdot \frac{\sum_{r=a,c} P_{Yk,r} - \delta}{\sum_{r=b,d} P_{Yk,r} + \delta} \quad (17)$$

where the extreme points are respectively

$$\delta^{X^{(1)}*} = \frac{\sum_{r=c,d} P_{Yj,r} \sum_{r=a,b} P_{Yk,r} - \sum_{r=a,b} P_{Yj,r} \sum_{r=c,d} P_{Yk,r}}{\sum_{\substack{i=j,k \\ r \in \mathcal{P}}} P_{Yi,r}} \quad (18)$$

$$\delta^{X^{(2)}*} = \frac{\sum\limits_{r=b,d} P_{Yj,r} \sum\limits_{r=a,c} P_{Yk,r} - \sum\limits_{r=a,c} P_{Yj,r} \sum\limits_{r=b,d} P_{Yk,r}}{\sum\limits_{\substack{i=j,k \\ r \in \mathcal{P}}} P_{Yi,r}} \quad (19)$$

Because the second-order derivatives of $\Delta H_{jk}^{X^{(1)}}$ and $\Delta H_{jk}^{X^{(2)}}$ with respect to $\delta$ are negative in the entire domain of $\delta$, both derivatives of $\Delta H_{jk}^{X^{(1)}}$ and $\Delta H_{jk}^{X^{(2)}}$ with respect to $\delta$ have at most one zero point, which means that both $\Delta H_{jk}^{X^{(1)}}$ and $\Delta H_{jk}^{X^{(2)}}$ are concave functions that cross the origin and have at most one maximum point.

$$\frac{d^2 \Delta H_{jk}^{X^{(1)}}}{d\delta^2} = -\frac{1}{\sum\limits_{r=c,d} P_{Yj,r} - \delta} - \frac{1}{\sum\limits_{r=a,b} P_{Yj,r} + \delta}$$
$$-\frac{1}{\sum\limits_{r=a,b} P_{Yk,r} - \delta} - \frac{1}{\sum\limits_{r=c,d} P_{Yk,r} + \delta} < 0 \quad (20)$$

$$\frac{d^2 \Delta H_{jk}^{X^{(2)}}}{d\delta^2} = -\frac{1}{\sum\limits_{r=b,d} P_{Yj,r} - \delta} - \frac{1}{\sum\limits_{r=a,c} P_{Yj,r} + \delta}$$
$$-\frac{1}{\sum\limits_{r=a,c} P_{Yk,r} - \delta} - \frac{1}{\sum\limits_{r=b,d} P_{Yk,r} + \delta} < 0 \quad (21)$$

The propositions about the mutual information pair are based on two basic assumptions, ensuring that the joint distribution of private characteristic $X^{(1)}$ and nonprivate characteristic $X^{(2)}$ must have four possible values, and that the unshared data $Y$ and shared data $U$ must have at least two possible intervals to meet the needs of practical scenarios. Under the basic assumptions, there are three propositions regarding the upper or lower bounds of the mutual information pair in the preventive audit.

***Proposition 1:*** The mutual information of the shared data $U$ and the private characteristic $X^{(1)}$ reaches a minimum value of 0, if and only if, for $\forall i \in \mathcal{Y}$, the shared data $U$ meet the condition:

$$P_{Ui,a} = P_{Ui,b} = P_{Ui,c} = P_{Ui,d} = 0 \vee \frac{P_{Ui,a} + P_{Ui,b}}{P_{Ui,c} + P_{Ui,d}} = \frac{p_1}{1-p_1} \quad (22)$$

***Proof:*** Refer to Appendix A.

***Proposition 2:*** The mutual information of the shared data $U$ and the nonprivate characteristic $X^{(2)}$ reaches a maximum value of $H(X^{(2)})$, if and only if, for $\forall i \in \mathcal{Y}$, the shared data $U$ meet the condition:

$$P_{Ui,a} + P_{Ui,c} = 0 \vee P_{Ui,b} + P_{Ui,d} = 0 \quad (23)$$

***Proof:*** Refer to Appendix B.

***Proposition 3:*** A decrease in the mutual information of the shared data $U$ and the private characteristic $X^{(1)}$, and an increase in the mutual information of the shared data $U$ and the nonprivate characteristic $X^{(2)}$ cannot be achieved simultaneously; that is, the sufficient condition for the mutual information pair to reach the Pareto optimal bound is that the shared data $U$ satisfy one of Eqs. (24) and (25) and Eq. (26).

$$\left| \sum_{r=b,d} P_{Uk,r} - \sum_{r=a,c} P_{Uj,r} \right| \geq \sum_{r=b,d} P_{Uj,r}, \sum_{r=a,c} P_{Uk,r} = 0, \sum_{r=b,d} P_{Uk,r} \neq 0 \quad (24)$$

$$\left| \sum_{r=a,c} P_{Uk,r} - \sum_{r=b,d} P_{Uj,r} \right| \geq \sum_{r=a,c} P_{Uj,r}, \sum_{r=a,c} P_{Uk,r} \neq 0, \sum_{r=b,d} P_{Uk,r} = 0 \quad (25)$$

$$\mathcal{Y} = \mathcal{Y}_1 + \mathcal{Y}_2 + \mathcal{Y}_3 + \mathcal{Y}_4 \quad (26)$$

where

$$\mathcal{Y}_1 = \left\{ i \in \mathcal{Y} \mid P_{Ui,r} = 0, \forall r \in \mathcal{P} \right\} \quad (27)$$

$$\mathcal{Y}_2 = \left\{ i \in \mathcal{Y} \mid \sum_{r=a,c} P_{Ui,r} \neq 0, \sum_{r=b,d} P_{Ui,r} = 0 \right\} \quad (28)$$

$$\mathcal{Y}_3 = \left\{ i \in \mathcal{Y} \mid \sum_{r=a,c} P_{Ui,r} = 0, \sum_{r=b,d} P_{Ui,r} \neq 0 \right\} \quad (29)$$

$$\mathcal{Y}_4 = \left\{ i,j \in \mathcal{Y} \left| \begin{array}{l} \sum\limits_{r=a,c} P_{Ui,r} \sum\limits_{r=b,d} P_{Ui,r} \neq 0, \\ \sum\limits_{r=a,c} P_{Uj,r} \sum\limits_{r=b,d} P_{Uj,r} \neq 0, \\ \sum\limits_{r=c,d} P_{Ui,r} \sum\limits_{r=a,b} P_{Uj,r} = \sum\limits_{r=c,d} P_{Uj,r} \sum\limits_{r=a,b} P_{Ui,r} \end{array} \right. \right\} \quad (30)$$

***Proof:*** Refer to Appendix C.

### B. Probability Exchange Adjustment Method with Variable Probability Distributions

According to the expression of mutual information, if the probability distribution of shared data $U$ and the probability distribution of unshared data $Y$ are different, the joint distribution of private characteristic $X^{(1)}$ and nonprivate characteristic $X^{(2)}$ should remain unchanged, but the information entropy of unshared data $Y$ can change. The mutual information of unshared data $Y$ and private characteristic $X^{(1)}$ can be decreased, and the mutual information of unshared data $Y$ and nonprivate characteristic $X^{(2)}$ can be increased. Therefore, preventive audits can achieve the above objectives by adjusting the probability distribution of $Y_{i,r}$ as shown below:

$$P_{Uj,r} - P_{Yj,r} = P_{Yk,r} - P_{Uk,r} = \delta \quad (31)$$

where $j, k \in \mathcal{Y}, r \in \mathcal{P}$, $\delta$ is the probability adjustment value.

In the probability exchange adjustment method with a variable probability distribution, when $r$ is $a$, $b$, $c$, or $d$, the mutual information of unshared data $Y$ and private characteristic $X^{(1)}$, as well as the mutual information of unshared data $Y$ and nonprivate characteristic $X^{(2)}$, can be changed simultaneously. For example, when $r$ is $a$, the change in the mutual information pair is shown in Fig. 10.

When $r$ is $a$, the information entropy of shared data $U$, the mutual information of shared data $U$ and private characteristic $X^{(1)}$, and the mutual information of shared data $U$ and nonprivate characteristic $X^{(2)}$ will change; thus, the change in the condition information entropy of shared data $U$ conditioned on private characteristic $X^{(1)}$ is no longer used as an equivalent representation of the change in the mutual information of shared data $U$ and private characteristic $X^{(1)}$. Similarly, the change in the conditional information entropy of shared data $U$ conditioned on nonprivate characteristic $X^{(2)}$ is no longer used as an equivalent representation of the change in the mutual information of shared data $U$ and nonprivate characteristic $X^{(2)}$. Instead, the changes in the mutual information of shared data $U$ and private characteristic $X^{(1)}$ and nonprivate characteristic $X^{(2)}$ are directly calculated, as shown below:

$$\Delta I_{jk}^{X^{(1)}}(\delta) = \Delta I_{aa,jk}^{X^{(1)}} = \Delta I_{bb,jk}^{X^{(1)}}$$
$$= \sum_{r=a,b} P_{Yj,r} \log \frac{\sum_{r=a,b} P_{Yj,r}}{\sum_{r=a,b} P_{Yj,r} + \delta} + \sum_{r=a,b} P_{Yk,r} \log \frac{\sum_{r=a,b} P_{Yk,r}}{\sum_{r=a,b} P_{Yk,r} - \delta}$$
$$+ \sum_{r \in \mathcal{P}} P_{Yj,r} \log \frac{\sum_{r \in \mathcal{P}} P_{Yj,r}}{\sum_{r \in \mathcal{P}} P_{Yj,r} + \delta} + \sum_{r \in \mathcal{P}} P_{Yk,r} \log \frac{\sum_{r \in \mathcal{P}} P_{Yk,r}}{\sum_{r \in \mathcal{P}} P_{Yk,r} - \delta} \quad (32)$$
$$+ \delta \log \frac{\sum_{r=a,b} P_{Yk,r} - \delta}{\sum_{r=a,b} P_{Yj,r} + \delta} \frac{\sum_{r \in \mathcal{P}} P_{Yk,r} - \delta}{\sum_{r \in \mathcal{P}} P_{Yj,r} + \delta}$$

$$\Delta I_{jk}^{X^{(2)}}(\delta) = \Delta I_{aa,jk}^{X^{(1)}} = \Delta I_{cc,jk}^{X^{(1)}}$$
$$= \sum_{r=a,c} P_{Yj,r} \log \frac{\sum_{r=a,c} P_{Yj,r}}{\sum_{r=a,c} P_{Yj,r} + \delta} + \sum_{r=a,c} P_{Yk,r} \log \frac{\sum_{r=a,c} P_{Yk,r}}{\sum_{r=a,c} P_{Yk,r} - \delta}$$
$$+ \sum_{r \in \mathcal{P}} P_{Yj,r} \log \frac{\sum_{r \in \mathcal{P}} P_{Yj,r}}{\sum_{r \in \mathcal{P}} P_{Yj,r} + \delta} + \sum_{r \in \mathcal{P}} P_{Yk,r} \log \frac{\sum_{r \in \mathcal{P}} P_{Yk,r}}{\sum_{r \in \mathcal{P}} P_{Yk,r} - \delta} \quad (33)$$
$$+ \delta \log \frac{\sum_{r=a,c} P_{Yk,r} - \delta}{\sum_{r=a,c} P_{Yj,r} + \delta} \frac{\sum_{r \in \mathcal{P}} P_{Yk,r} - \delta}{\sum_{r \in \mathcal{P}} P_{Yj,r} + \delta}$$

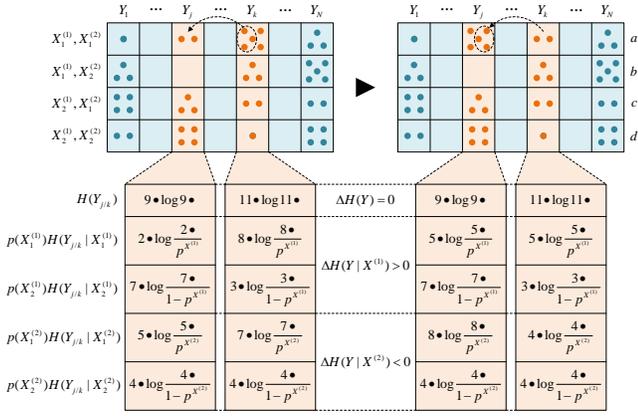

Fig. 10. Inversely changed mutual information pair after probability exchange adjustment method with variable probability distribution.

## IV. OPTIMIZATION MODEL OF PREVENTIVE AUDIT AND ITS EXTENSIONS

For unshared data $Y$, shared data $U$, private characteristic $X^{(1)}$, and nonprivate characteristic $X^{(2)}$, the complete optimization model of preventive audits for data applications is as follows:

$$\min_{P_{Ui,r}, i \in \mathcal{Y}, r \in \mathcal{P}} \lambda_1 I(X^{(1)}; U) + \lambda_2 I(X^{(2)}; U) \quad (34)$$

$$s.t. \quad I(X^{(1)}; U) \leq I_{set}^{X^{(1)}} \quad (35)$$

$$I(X^{(2)}; U) \geq I_{set}^{X^{(2)}} \quad (36)$$

$$\sum_r P_{Ui,r} = \sum_r P_{Yi,r}, \forall i \in \mathcal{Y} \quad (37)$$

$$\sum_{i=1}^{N} P_{Ui,r} = \sum_{i=1}^{N} P_{Yi,r}, \forall r \in \mathcal{P} \quad (38)$$

where $\lambda_1$ and $\lambda_2$ is the weight coefficient of the mutual information of shared data $U$ and private characteristic $X^{(1)}$ in the optimization objective, and the weight coefficient of mutual information of shared data $U$ and nonprivate characteristic $X^{(2)}$, respectively. The upper bound of the privacy index is set by the data owners based on the mutual information of shared data $U$ and private characteristic $X^{(1)}$, and the lower bound of utility index is set by the data service providers based on the mutual information of shared data $U$ and nonprivate characteristic $X^{(2)}$.

According to the classification of the privacy-utility trade-offs in Section III-C, when data owners are dominant, the weight coefficient $\lambda_1=0$ and the weight coefficient $\lambda_2$ can be any negative number, and the utility requirement in Eq. (36) can be ignored; that is, the preventive audit maximizes the utility index of the data service providers under their privacy requirements. When data service providers are dominant, the weight coefficient $\lambda_2=0$ and the weight coefficient $\lambda_1$ can be any positive number, and the privacy requirement in Eq. (35) can be ignored, that is, the preventive audit minimizes the privacy index of data owners under utility requirements of data service providers. When the market power of the data owners and data service providers are similar, the weight coefficient $\lambda_1$ can be any positive number, while the weight coefficient $\lambda_2$ can be any negative number, and the ratio between $\lambda_1$ and $\lambda_2$ determines the privacy-utility trade-off.

In addition, Eq. (37) is optional, which indicates that the probability distribution of shared data $U$ and the probability distribution of unshared data $Y$ are the same, ensuring that the information entropy of shared data $U$ and the information entropy of unshared data $Y$ are equal. If the preventive audit requires that the probability distribution of shared data $U$ and the probability distribution of unshared data $Y$ be the same to avoid affecting the form of unshared data as much as possible, the constraint needs to be set. Eq. (38) is mandatory, which indicates that the joint distribution of the private characteristic $X^{(1)}$ and nonprivate characteristic $X^{(2)}$ remains unchanged. Since private characteristic $X^{(1)}$ and nonprivate characteristic $X^{(2)}$ will not change through preventive audits, the joint distribution of private characteristic $X^{(1)}$ and nonprivate characteristic $X^{(2)}$ should remain unchanged.

By simplifying both private characteristic $X^{(1)}$ and nonprivate characteristic $X^{(2)}$ into binary discrete random variables, the probability exchange adjustment method and propositions about the mutual information pair on the optimal bound can be used to gradually approach the optimal bound.

Considering that private characteristic $X^{(1)}$ and nonprivate characteristic $X^{(2)}$ may not necessarily have only two values in practical scenarios, the different value combinations of private characteristic $X^{(1)}$ and nonprivate characteristic $X^{(2)}$ may far exceed 4, but the values of the mutual information pairs can still be changed through the probability exchange adjustment method. In addition, the number of private and nonprivate characteristics can also increase. The diverse privacy requirements of data owners may lead to multiple private characteristics $X^{(u)}$, such as the end users' gender, employment situation, education level, and social class, while the diverse utility requirements of data service providers may lead to multiple nonprivate characteristics $X^{(v)}$, such as the proportion of light bulbs and cook type.

Note that there is mutual information of each private characteristic $X^{(u)}$, $u \in \mathcal{D}^u$ and each nonprivate characteristic $X^{(v)}$, $v \in \mathcal{D}^v$. Therefore, in the process of privacy-utility trade-offs, it cannot be guaranteed that the mutual information of shared data $U$ and each private characteristic $X^{(u)}$ reaches the minimum possible value, while the mutual information of shared data $U$ and each nonprivate characteristic $X^{(v)}$ also reaches its maximum possible value. Due to the coupling between the mutual information pairs, if there are multiple private characteristics $X^{(u)}$ and multiple nonprivate characteristics $X^{(v)}$, when the mutual information indices of shared data $U$ and several private characteristics $X^{(u)}$ are reduced to a minimum value of 0, and the mutual information of shared data $U$ and other private characteristics $X^{(u)}$ remains unchanged, there is an upper bound on the mutual information of shared data $U$ and any nonprivate characteristic $X^{(v)}$:

$$I(X^{(v)};U) \leq H(X^{(v)} \mid X^{(u_1)}, X^{(u_2)}, \cdots, X^{(u_{U_c})}), U_c = \left|\mathcal{D}_c^u\right| \quad (39)$$

Because the upper bound of the mutual information of shared data $U$ and any nonprivate characteristic $X^{(v)}$ should not be lower than the mutual information of unshared data $Y$ and the nonprivate characteristic $X^{(v)}$, the shared data $U$ obtained by data service providers can still be used to infer and use the nonprivate characteristic $X^{(v)}$ without being affected after being processed by the preventive audit. Therefore, the interests of both data sharing entities, that is, the data owners and data service providers, are fully considered.

The extended optimization model is as follows:

$$\min_{P_{Ui,r}, i \in \mathcal{Y}, r \in \mathcal{P}} \sum_{u \in \mathcal{D}^u} \lambda_{1,u} I(X^{(u)};U) + \sum_{v \in \mathcal{D}^v} \lambda_{2,v} I(X^{(v)};U) \quad (40)$$

$$\text{s.t.} \quad I(X^{(u)};U) \leq I_{set}^{X^{(u)}}, \forall u \in \mathcal{D}^u \quad (41)$$

$$I(X^{(v)};U) \geq I_{set}^{X^{(v)}}, \forall v \in \mathcal{D}^v \quad (42)$$

$$\sum_r P_{Ui,r} = \sum_r P_{Yi,r}, \forall i \in \mathcal{Y} \quad (43)$$

$$\sum_{i=1}^N P_{Ui,r} = \sum_{i=1}^N P_{Yi,r}, \forall r \in \mathcal{P} \quad (44)$$

V. CASE STUDIES

A. Basic Setup

In the case studies, the smart meter data of electricity consumption provided by the CER in Ireland are selected as electric load data, while the end users' social characteristic data are derived from the results of the residential pre-trial survey issued by the CER. Overall, 143 kinds of social characteristics of the end users are counted in the results, and 10 kinds of social characteristics that are representative of, or are more relevant to, the electric load data are selected as typical social characteristics of the end users. The 10 social characteristics are shown in Table I, and they are classified as end users' private or nonprivate characteristics according to the requirements of case studies. In addition, the 10 social characteristics included 5 social characteristics that are representative, namely, employment status, social class, home type, own or rent, level of education, and 5 social characteristics that are more related to the electric load data, namely, internet access, family members, number of bedrooms, cook type, and proportion of light bulbs.

B. Probability Exchange Adjustment by Step

Firstly, to comply with the assumption that all social characteristics are treated as binary discrete random variables, the aforementioned 10 social characteristics should be modified from multivariate discrete random variables to binary discrete random variables. The method of modification involves separating the values of the social characteristics into two parts that are close to each other internally. For example, for the 4th social characteristic "family members", the first and second options can be regarded as one part, the third option can be regarded as the other part, or the first option can be regarded as one part, the second and third options can be regarded as the other part. Among the different choices of partitioning methods, the partitioning method that maximizes the mutual information of the electric load data and each social characteristic is adopted. Based on the mutual information of the electric load data and 10 binary discrete random variables of the social characteristics, the social characteristics with the highest two mutual information are selected as private characteristic $X^{(1)}$ and nonprivate characteristic $X^{(2)}$ to increase the significance of the preventive audit. In this section, "internet access" is regarded as a private characteristic $X^{(1)}$, "family members" is regarded as a nonprivate characteristic $X^{(2)}$, the original and audited electric load data are unshared data $Y$ and shared data $U$, respectively.

TABLE I
10 SOCIAL CHARACTERISTICS OF END USERS AND THEIR MAIN OPTIONS

| # | Social characteristics | Main options |
|---|---|---|
| 1 | Employment status | An employee/Self-employed/Unemployed/Retired/Carer: Looking after relative family |
| 2 | Social class | AB/C1/C2/DE/F [RECORD ALL FARMERS] |
| 3 | Internet access | Yes/No |
| 4 | Family members | I live alone/All people in my home are over 15 years of age/Both adults and children under 15 years of age live in my home |
| 5 | Home type | Apartment/Semi-detached house/Detached house/Terraced house/Bungalow |
| 6 | Own or rent | Rent (from a private landlord)/Rent (from a local authority)/Own Outright (not mortgaged)/Own with mortgage etc/Other |
| 7 | Number of bedrooms | 1/2/3/4/5+ |
| 8 | Cook type | Electric cooker/Gas cooker/Oil fired cooker/Solid fuel cooker (stove aga) |
| 9 | Proportion of light bulbs | None/About a quarter/About half/About three quarters/All |
| 10 | Level of education | No formal education/Primary/Secondary to Intermediate Cert Junior Cert level/Secondary to Leaving Cert level/Third level |

According to the introduction in Section IV-A, the change of the mutual information of the electric load data and the two social characteristics can be realized through probability exchange adjustment methods under the condition that the probability distribution of the electric load data remains unchanged and the joint distribution of "internet access" and "family members" remains unchanged:

a) Step 1: Change $P_{Yj,r}$, $P_{Yj,s}$, $P_{Yk,r}$, $P_{Yk,s}$ corresponding to "internet access" and "family members" through an adjustment of the original electric load data without changing $P_{Yj}$ and $P_{Yk}$, where $r$ and $s$ are $a$ and $d$, $d$ and $a$, $b$ and $c$, $c$ and $b$, while simultaneously decreasing the mutual information of the electric load data and "internet access" and increasing the mutual information of the electric load data and "family members";

b) Step 2: Change $P_{Uj,r}$, $P_{Uj,s}$, $P_{Uk,r}$, $P_{Uk,s}$ corresponding to "internet access" and "family members" without changing $P_{Uj}$ and $P_{Uk}$ for the audited electric load data after Step 1, where $r$ and $s$ are $a$ and $c$, $c$ and $a$, $b$ and $d$, $d$ and $b$, and decrease the mutual information of the electric load data and "internet access";

c) Step 3: Change $P_{Uj,r}$, $P_{Uj,s}$, $P_{Uk,r}$, $P_{Uk,s}$ corresponding to "internet access" and "family members" without changing $P_{Uj}$ and $P_{Uk}$ for the audited electric load data after Step 2, where $r$ and $s$ are $a$ and $b$, $b$ and $a$, $c$ and $d$, $d$ and $c$, and increase the mutual information of the electric load data and "family members".

Through the preventive audit in Steps 1, 2, and 3, the changes in the probability distribution of the electric load data by interval are shown in Fig. 11, and the changes in the mutual information pair are shown in Fig. 12.

Fig. 11 shows that the probability distribution of the electric load data remains the same before and after the preventive audit. Although adjustments have been made to the electric load data through the probability exchange adjustment method, by changing the probability distribution of the electric load data conditioned on the end users with different social characteristics, the overall probability distribution of the electric load data has not changed. Therefore, after the adjustments in Steps 1, 2, and 3, the internal composition of $P_{Uj}$ varies.

Fig. 12 shows that the initial value of the mutual information pair is (0.0180, 0.0294); after the adjustment in Step 1, the mutual information pair changes to (0.0021, 0.1536); after the adjustment in Step 2, the mutual information pair changes to (0.0002, 0.1536); and after the adjustment in Step 3, the mutual information pair changes to (0.0002, 0.2571). The change process of the mutual information pair conforms to the theoretical results. It is known that when the mutual information of the electric load data and "internet access" decreases, it is difficult for data service providers to infer whether end users have internet access by using the shared electric load data, otherwise, data service providers may infer information about end users' Internet access and sell the information to broadband agents to earn additional data value, which may not be allowed by the data owners. At the same time, when the mutual information of the electric load data and "family members" is not decreased, the data service providers can infer end users' family members to at least the same extent based on the obtained electric load data before and after the preventive audits. Undoubtedly, end users' family members help the data service providers better forecast end users' future electric load and serve the electricity retailers.

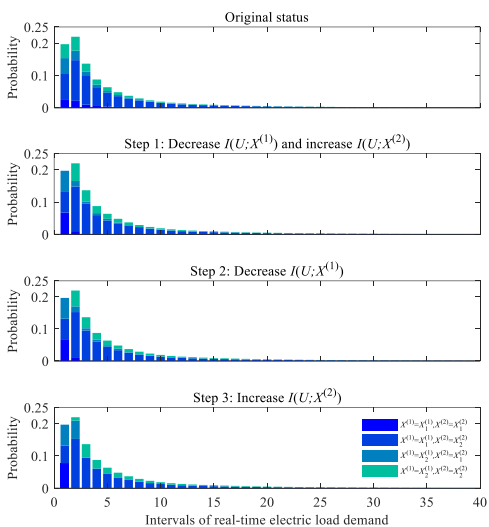

Fig. 11. Change of the probability distribution of electric load data by step through the probability exchange adjustment method with constant probability distribution.

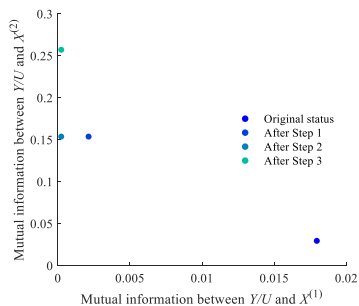

Fig. 12. Change of the mutual information pair by step through the probability exchange adjustment method with constant probability distribution.

### C. Probability Exchange Adjustment by Optimization Model

Based on the optimization model of the preventive audit, the original electric load data and audited electric load data are still regarded as unshared data $Y$ and shared data $U$, and "internet access" and "family members" are still regarded as private characteristic $X^{(1)}$ and nonprivate characteristic $X^{(2)}$ respectively.

For the case where the probability distribution is constant, in the objective function, the weight coefficient of the mutual information of the electric load data and "internet access" $\lambda_1$ is -1, and the weight coefficient of the mutual information of the electric load data and "family members" $\lambda_2$ is 1. Among the constraints, the constraint that the probability distribution of the electric load data remains unchanged, the constraint that the joint distribution of "internet access" and "family members" remains unchanged, the lower bound constraint of the mutual information of the electric load data and "Internet access" are all retained, while the upper bound constraint of the mutual information of the electric load data and "family members" is ignored. Besides, the lower bound of the mutual information of the electric load data and "internet access" is set to have 25 possible values from 0.002 to 0.05 with an interval of 0.002.

When the lower bounds of the mutual information of the audited electric load data and "internet access" meet the 25 set values respectively, the optimal probability distribution of the electric load data under the privacy-utility trade-off can be obtained by solving the optimization model of the preventive audit. Then, when the lower bound is at the 3rd, 12th, and 20th set values, the changes in the probability distribution of the electric load data by interval are shown in Fig. 13, and the changes in the mutual information pair are shown in Fig. 14.

Fig. 13 shows that the probability distribution of audited electric load data always remains the same as the probability distribution of the original electric load data. Although the electric load data of the end users with different social characteristics are directly adjusted according to the lower bound constraints; changing the probability distribution of the electric load data conditioned on end users with different social characteristics, the overall probability distribution of the electric load data has not changed. Therefore, when the lower bound of the mutual information of the electric load data and "internet access" meet the 3rd, 12th, and 20th set values, the internal composition of $P_{Uj}$ varies.

Fig. 14 shows that the mutual information pair are less than the theoretical optimal bound, and the envelope line of most of the mutual information pairs is the Pareto optimal bound. On the Pareto optimal bound, through the probability exchange adjustment method with constant probability distribution, it is impossible to unilaterally increase the mutual information of the electric load data and "family members" without changing the mutual information of the electric load data and "internet access", or unilaterally decrease the mutual information of the electric load data and "internet access" without changing the mutual information of the electric load data and "family members". By fitting a polynomial function, a quadratic function curve can be obtained, which is basically in line with the envelope line of the mutual information pairs, and can be used to accurately depict the Pareto optimal bound.

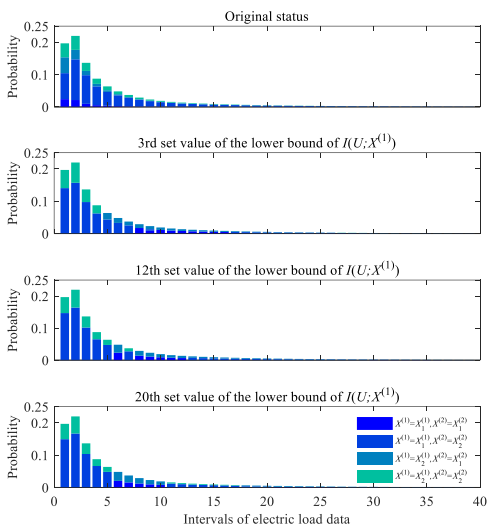

Fig. 13. Change of the probability distribution of electric load data at different set values of the lower bound through the probability exchange adjustment method with constant probability distribution.

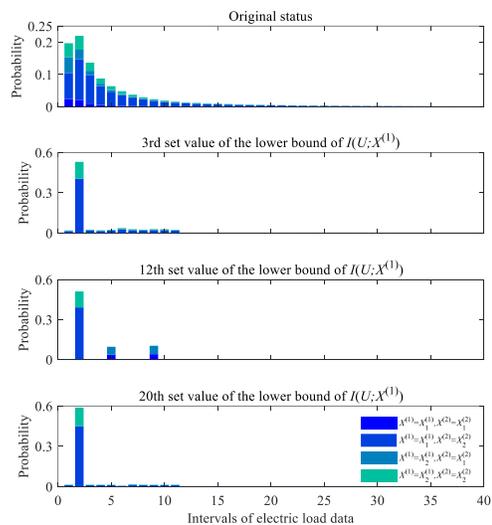

Fig. 15. Change of the probability distribution of electric load data at different set values of the lower bound through the probability exchange adjustment method with variable probability distribution.

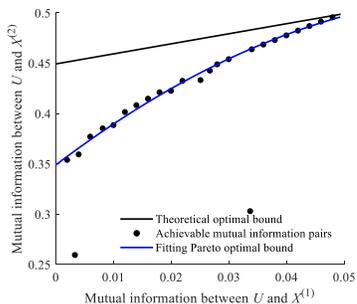

Fig. 14. Change of the mutual information pair at different set values of the lower bound through the probability exchange adjustment method with constant probability distribution.

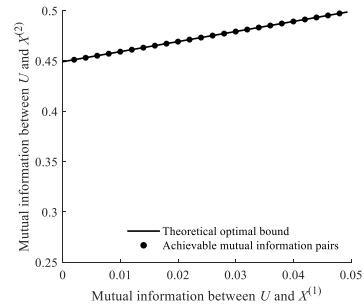

Fig. 16. Change of the mutual information pair at different set values of the lower bound through the probability exchange adjustment method with variable probability distribution.

For the case where the probability distribution is variable, the objective function and most of the constraints remain unchanged, while the constraint that the probability distribution of the electric load data remains unchanged is ignored. In addition, the lower bound is still set to have 25 possible values from 0.002 to 0.05 with an interval of 0.002. For the 3rd, 12th, and 20th set values of the lower bound, the changes in the probability distribution of the electric load data by interval are shown in Fig. 15, and the changes in the mutual information pair are shown in Fig. 16.

Fig. 15 shows that the probability distribution of electric load data significantly changes, before and after, the preventive audit. When the lower bound of the mutual information of the electric load data and "internet access" meets different set values, the probability distribution of the electric load data will be different, with the probability in the second interval approaching 0.6, while the probability in the other intervals is relatively low.

Fig. 16 shows that the mutual information pairs are all located on the theoretical optimal bound. On the theoretical optimal bound, through the probability exchange adjustment method with variable probability distribution, it is impossible to unilaterally increase the mutual information of the electric load data and "family members" without changing the mutual information of the electric load data and "internet access", or unilaterally decrease the mutual information of the electric load data and "internet access" without changing the mutual information of the electric load data and "family members". Therefore, the probability exchange adjustment method with a variable probability distribution can achieve a theoretical optimal bound rather than being limited by the Pareto optimal bound. In addition, on the theoretical optimal bound, the data service providers cannot more accurately simultaneously infer whether the end users have internet access and the end users' family members, by analyzing the electric load data after the preventive audit. If the data service providers set higher utility requirements, the data owners will lose some of the privacy requirements to meet the utility requirements of the data service providers. However, the privacy and utility requirements cannot be met simultaneously, which may make it difficult to achieve data sharing between the data owners and data service providers. Therefore, in preventive audits, it is necessary to avoid the mutual information pair being located on the theoretical optimal bound.

### D. Probability Exchange Adjustment by Extended Optimization Model

Based on the extended optimization model of preventive audits, the original electric load data and audited electric load data are still regarded as unshared data $Y$ and shared data $U$, "internet access" is still regarded as private characteristic $X^{(1)}$, "family members" and "number of bedrooms" are regarded as private characteristic $X^{(2)}$ and nonprivate characteristic $X^{(3)}$, respectively. To better match the practical scenarios, "internet access" and "number of bedrooms" are reserved as binary discrete random variables, and "family members" is a ternary discrete random variable. Therefore, there are 12 possible values for the group of three social characteristics.

For the case where the probability distribution is constant, in the objective function, the weight coefficient of the mutual information of the electric load data and "internet access" $\lambda_1$ is -1, the weight coefficient of the mutual information of the electric load data and "family members" $\lambda_2$ is -1, and the weight coefficient of the mutual information of the electric load data and "number of bedrooms" $\lambda_3$ is 1. In addition, the lower bound of the mutual information of the electric load data and "internet access" is set to have 6 possible values from 0 to 0.025 with an interval of 0.005, while the lower bound of the mutual information of the electric load data and "family members" is set to have 7 possible values from 0 to 0.03 with an interval of 0.005.

When the lower bound of the mutual information of the audited electric load data and "internet access" and the mutual information of the audited electric load data and "family members" meet the 35 set values, the optimal probability distribution of the electric load data under the privacy-utility trade-off can be obtained by solving the optimization model of the preventive audit. The changes in mutual information groups formed by the mutual information of the electric load data and "internet access", "family members" and "number of bedrooms" are shown in Fig. 17. For the case where the probability distribution is variable, the changes in the mutual information groups are shown in Fig. 18.

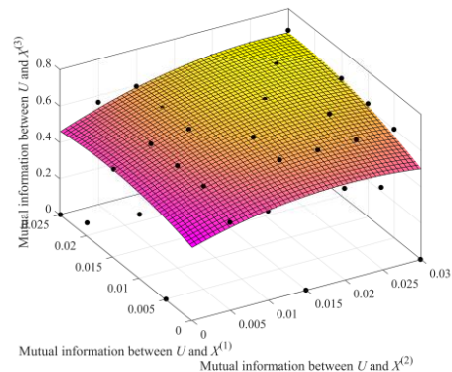

Fig. 17. Change of the mutual information group at different set values of the lower bound through the probability exchange adjustment method with constant probability distribution.

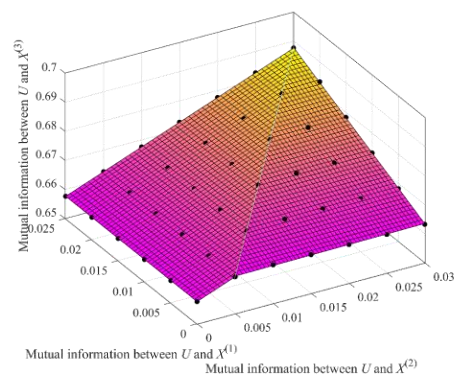

Fig. 18. Change of the mutual information group at different set values of the lower bound through the probability exchange adjustment method with variable probability distribution.

Fig. 17 shows that the mutual information groups are lower than the three-dimensional theoretical optimal plane. The envelope surface of most of the mutual information groups is the Pareto optimal plane of the mutual information, and its physical meaning is close to that of the two-dimensional Pareto optimal bound. Similarly, by fitting a polynomial function, a surface can be obtained that is basically in line with the envelope surface of the mutual information groups, and can be used to accurately depict the Pareto optimal plane. Fig. 18 shows that the mutual information groups are all located on the three-dimensional theoretical optimal plane.

Therefore, the probability exchange adjustment methods with constant and variable probability distributions can both be extended to those with multivariate social characteristics.

## VI. CONCLUSIONS AND FUTURE WORK

Focusing on the data sharing in the power IoT, the prevent audits for data applications are proposed to achieve privacy-utility trade-offs in this paper. Firstly, from the perspective of semantic information theory, the semantic mutual information is selected as the privacy and utility indices based on interpretability and generality, and then different types of preventive audits can be formed based on the privacy and utility requirements. Secondly, probability exchange adjustment methods with constant and variable probability distributions are proposed for different demands, and corresponding propositions are proposed and proven to determine the limits of the probability exchange adjustment methods. Thirdly, step-by-step probability exchange adjustment methods can be abstracted and summarized as the optimization model of the preventive audit, and this approach can be extended for the multivariable social characteristics of data owners. Finally, the results of the case studies validate that the mutual information pair can be inversely changed through a reasonable design of the probability exchange adjustment methods and that the mutual information pair will reach its theoretical or Pareto optimal bound corresponding to the optimal solutions. These optimal bounds are in line with the propositions, which may provide the guidance of preventive audits to entities in data sharing.

Future works can be extended to the design of data transaction systems. The support of multiple methods, such as federated learning, is expected to form an institutional foundation for data transaction systems through the design of data clearing mechanisms, smart contracts, and regulatory systems.

## APPENDIX

### A. Proof of Proposition 1

Firstly, consider the sufficiency of the condition.
According to Eq. (22), we have

$$\sum_{r=c,d} P_{Uj,r} \sum_{r=a,b} P_{Uk,r} - \sum_{r=c,d} P_{Uk,r} \sum_{r=a,b} P_{Uj,r} = 0 \quad (A.1)$$

The derivative of $\Delta H_{jk}^{X^{(1)}}$ with respect to $\delta$ for $\forall j,k \in \mathcal{Y}$ can be simplified as follows:

$$\frac{d\Delta H_{jk}^{X^{(1)}}(\delta)}{d\delta} = \log \frac{\sum_{r=c,d} P_{Yj,r} - \delta}{\sum_{r=a,b} P_{Yj,r} + \delta} \frac{\sum_{r=a,b} P_{Yk,r} - \delta}{\sum_{r=c,d} P_{Yk,r} + \delta} \begin{cases} > 0, \delta < 0 \\ = 0, \delta = 0 \\ < 0, \delta > 0 \end{cases} \quad (A.2)$$

Therefore, $\Delta H_{jk}^{X^{(1)}}$ only have one extreme point $\delta=0$, and reaches the maximum value at the extreme point:

$$\Delta H_{jk}^{X^{(1)}}(\delta) \leq 0, \forall j,k,\delta \quad (A.3)$$

Because the information entropy of shared data $U$ is locked in $H(U)$, when the conditional information entropy of shared data $U$ conditioned on private characteristic $X^{(1)}$ cannot increase, the mutual information of shared data $U$ and private characteristic $X^{(1)}$ cannot decrease.

Then, we define set $\mathcal{Y}_5$ as

$$\mathcal{Y}_5 = \left\{ i \in \mathcal{Y} \middle| (1-p_1) \sum_{r=a,b} P_{Ui,r} = p_1 \sum_{r=a,b} P_{Ui,r} \right\} \quad (A.4)$$

When shared data $U$ satisfy Eq. (22), the conditional information entropy of shared data U conditioned on private characteristic $X^{(1)}$ equals to the information entropy of shared data $U$, that is, the mutual information of shared data $U$ and private characteristic $X^{(1)}$ is 0:

$$H(U \mid X^{(1)}) = -\sum_{\substack{i=1 \\ i \in \mathcal{Y}_5}}^{N} \left( \sum_{r=a,b} P_{Ui,r} \right) \log \frac{\sum_{r=a,b} P_{Ui,r}}{p_1} \\ -\sum_{\substack{i=1 \\ i \in \mathcal{Y}_5}}^{N} \left( \sum_{r=c,d} P_{Ui,r} \right) \log \frac{\sum_{r=c,d} P_{Ui,r}}{1-p_1} \quad (A.5)$$

Simplify the above equation, we have

$$H(U \mid X^{(1)}) = -p_1 \sum_{\substack{i=1 \\ i \in \mathcal{Y}_5}}^{N} P_{Ui} \log P_{Ui} - (1-p_1) \sum_{\substack{i=1 \\ i \in \mathcal{Y}_5}}^{N} P_{Ui} \log P_{Ui} \\ = -\sum_{\substack{i=1 \\ i \in \mathcal{Y}_5}}^{N} P_{Ui} \log P_{Ui} = -\sum_{i=1}^{N} P_{Ui} \log P_{Ui} = H(U) \quad (A.6)$$

Secondly, consider the necessity of the condition.
Using the reduction to absurdity, assuming that for $\exists i \in \mathcal{Y}$, the condition in Eq. (22) are not met, then for $\exists j,k \in \mathcal{Y}$, Eq. (A.1) must be satisfied, otherwise the solution to Eq. (A.1) will include $P_{Ui,a}+P_{Ui,b}=0$ or $P_{Ui,c}+P_{Ui,d}=0$, which contradicts the basic assumption.

$$\sum_{r=c,d} P_{Uj,r} \sum_{r=a,b} P_{Uk,r} \neq \sum_{r=c,d} P_{Uk,r} \sum_{r=a,b} P_{Uj,r} \quad (A.7)$$

Without loss of generality, it is assumed that for $\exists j,k \in \mathcal{Y}$,

$$\sum_{r=c,d} P_{Uj,r} \sum_{r=a,b} P_{Uk,r} < \sum_{r=c,d} P_{Uk,r} \sum_{r=a,b} P_{Uj,r} \quad (A.8)$$

Therefore, the extreme point is

$$\delta^{X^{(1)}*} = \frac{\sum_{r=c,d} P_{Uj,r} \sum_{r=a,b} P_{Uk,r} - \sum_{r=c,d} P_{Uk,r} \sum_{r=a,b} P_{Uj,r}}{\sum_{r\in\mathcal{P}} P_{Uj,r} + \sum_{r\in\mathcal{P}} P_{Uk,r}} < 0 \quad (A.9)$$

Besides, based on Eq. (A.9) and the non-negativity of $P_{Ui,r}$ for $\forall i \in \mathcal{Y}, \forall r \in \mathcal{P}$, we have

$$\sum_{r=a,b} P_{Uj,r} > 0, \sum_{r=c,d} P_{Uk,r} > 0 \quad (A.10)$$

According to the theoretical domain of $\delta$, the domain will at least retain the non-positive half. Because the derivative of $\Delta H_{jk}^{X^{(1)}}$ with respect to $\delta$ is negative on the interval $(\delta^{X^{(1)}*}, 0)$, thus $\Delta H_{jk}^{X^{(1)}}$ will be positive on the interval $(\delta^{X^{(1)}*}, 0)$, that is, the mutual information of shared data $U$ and private characteristic $X^{(1)}$ can still be further decreased, which contradicts the assumption that the mutual information reaches a minimum value of 0, then the assumption is incorrect. ∎

*B. Proof of Proposition 2*

Firstly, consider the sufficiency of the condition.
Based on the condition in Eq. (23), there exists three possible situations:

a) The first situation satisfies:

$$\sum_{r=b,d} P_{Uj,r} \sum_{r=a,c} P_{Uk,r} = \sum_{r=b,d} P_{Uk,r} \sum_{r=a,c} P_{Uj,r} = 0 \quad (B.1)$$

The extreme point is

$$\delta^{X^{(2)}*} = \frac{\sum_{r=b,d} P_{Uj,r} \sum_{r=a,c} P_{Uk,r} - \sum_{r=b,d} P_{Uk,r} \sum_{r=a,c} P_{Uj,r}}{\sum_{r\in\mathcal{P}} P_{Uj,r} + \sum_{r\in\mathcal{P}} P_{Uk,r}} = 0 \quad (B.2)$$

Therefore, the domain of $\delta$ only remains $\delta=0$, and the unique value of $\Delta H_{jk}^{X^{(2)}}$ is 0 at $\delta=0$.

b) The second situation satisfies:

$$\sum_{r=b,d} P_{Uj,r} \sum_{r=a,c} P_{Uk,r} \neq 0, \sum_{r=b,d} P_{Uk,r} \sum_{r=a,c} P_{Uj,r} = 0 \quad (B.3)$$

The extreme point is

$$\delta^{X^{(2)}*} = \frac{\sum_{r=b,d} P_{Uj,r} \sum_{r=a,c} P_{Uk,r} - \sum_{r=b,d} P_{Uk,r} \sum_{r=a,c} P_{Uj,r}}{\sum_{r\in\mathcal{P}} P_{Uj,r} + \sum_{r\in\mathcal{P}} P_{Uk,r}} > 0 \quad (B.4)$$

Therefore, the domain of $\delta$ only remains the non-negative half:

$$0 \leq \delta \leq \min\left(\sum_{r=a,c} P_{Uk,r}, \sum_{r=b,d} P_{Uj,r}\right) \quad (B.5)$$

Based on the symmetry, it can be assumed that

$$\sum_{r=a,c} P_{Uk,r} \leq \sum_{r=b,d} P_{Uj,r} \quad (B.6)$$

Therefore, the upper bound of $\delta$ is $P_{Uk,a}+P_{Uk,c}$, and the value of $\Delta H_{jk}^{X^{(2)}}$ when $\delta$ at its upper bound:

$$\Delta H_{jk}^{X^{(2)}}\left(\sum_{r=a,c} P_{Uk,r}\right)$$
$$= \sum_{r=b,d} P_{Uj,r} \log \sum_{r=b,d} P_{Uj,r} - \sum_{r=a,c} P_{Uk,r} \log \sum_{r=a,c} P_{Uk,r} \quad (B.7)$$
$$-\left(\sum_{r=b,d} P_{Uj,r} - \sum_{r=a,c} P_{Uk,r}\right) \log\left(\sum_{r=b,d} P_{Uj,r} - \sum_{r=a,c} P_{Uk,r}\right)$$

It can be easily proved that for $\forall x, y, x+y \in (0,1)$, the following inequality holds:

$$x \log x + y \log y < x \log(x+y) + y \log(x+y)$$
$$= (x+y)\log(x+y) \quad (B.8)$$

Therefore, the value of $\Delta H_{jk}^{X^{(2)}}$ in Eq. (B.7) is positive. Because the derivative of $\Delta H_{jk}^{X^{(2)}}$ with respect to $\delta$ is positive in the interval $[0, \delta^{X^{(2)}*}]$ and the derivative of $\Delta H_{jk}^{X^{(2)}}$ with respect to $\delta$ is negative in the interval $(\delta^{X^{(2)}*}, P_{Uk,a}+P_{Uk,c}]$, the value of $\Delta H_{jk}^{X^{(2)}}$ in the domain of $\delta$ will all be non-negative.

c) The third situation satisfies:

$$\sum_{r=b,d} P_{Uj,r} \sum_{r=a,c} P_{Uk,r} = 0, \sum_{r=b,d} P_{Uk,r} \sum_{r=a,c} P_{Uj,r} \neq 0 \quad (B.9)$$

The extreme point is

$$\delta^{X^{(2)}*} = \frac{\sum_{r=b,d} P_{Uj,r} \sum_{r=a,c} P_{Uk,r} - \sum_{r=b,d} P_{Uk,r} \sum_{r=a,c} P_{Uj,r}}{\sum_{r\in\mathcal{P}} P_{Uj,r} + \sum_{r\in\mathcal{P}} P_{Uk,r}} < 0 \quad (B.10)$$

Therefore, the domain of $\delta$ only remains the non-positive half:

$$-\min\left(\sum_{r=a,c} P_{Uj,r}, \sum_{r=b,d} P_{Uk,r}\right) \leq \delta \leq 0 \quad (B.11)$$

Based on the symmetry, it can be assumed that

$$\sum_{r=b,d} P_{Uk,r} \leq \sum_{r=a,c} P_{Uj,r} \quad (B.12)$$

Therefore, the lower bound of $\delta$ is $-P_{Uk,b}-P_{Uk,d}$, and the value of $\Delta H_{jk}^{X^{(2)}}$ when $\delta$ at its lower bound:

$$\Delta H_{jk}^{X^{(2)}}\left(-\sum_{r=b,d} P_{Uk,r}\right)$$
$$= \sum_{r=a,c} P_{Uj,r} \log \sum_{r=a,c} P_{Uj,r} - \sum_{r=b,d} P_{Uk,r} \log \sum_{r=b,d} P_{Uk,r} \quad (B.13)$$
$$-\left(\sum_{r=a,c} P_{Uj,r} - \sum_{r=b,d} P_{Uk,r}\right) \log\left(\sum_{r=a,c} P_{Uj,r} - \sum_{r=b,d} P_{Uk,r}\right)$$

According to Eq. (B.8), the value of $\Delta H_{jk}^{X^{(2)}}$ in Eq. (B.13) is positive. Because the derivative of $\Delta H_{jk}^{X^{(2)}}$ with respect to $\delta$ is positive in the interval $[-P_{Uk,b}-P_{Uk,d}, \delta^{X^{(2)}*}]$ and the derivative of $\Delta H_{jk}^{X^{(2)}}$ with respect to $\delta$ is negative in the interval $(\delta^{X^{(2)}*}, 0]$, the value of $\Delta H_{jk}^{X^{(2)}}$ in the domain of $\delta$ will all be non-negative.

Three possible situations have been discussed, and value of $\Delta H_{jk}^{X^{(2)}}$ in the domain of $\delta$ will not be negative.

Because the information entropy of shared data $U$ being locked to $H(U)$, when the conditional information entropy of shared data $U$ conditioned on nonprivate characteristic $X^{(2)}$ cannot decrease, the mutual information of shared data $U$ and non-privacy information characteristic $X^{(2)}$ cannot increase.

When shared data $U$ satisfies Eq. (23), the conditional information entropy of shared data $U$ conditioned on nonprivate characteristic $X^{(2)}$ is $H(U)-H(X^{(2)})$, that is, the mutual information of shared data $U$ and nonprivate characteristic $X^{(2)}$ is $H(X^{(2)})$.

$$H(U \mid X^{(2)}) = -\sum_{\substack{i=1\\i\in\mathcal{Y}_2}}^{N}\left(\sum_{r=a,c} P_{Ui,r}\right)\log\frac{\sum_{r=a,c} P_{Ui,r}}{p_2} \\ -\sum_{\substack{i=1\\i\in\mathcal{Y}_3}}^{N}\left(\sum_{r=b,d} P_{Ui,r}\right)\log\frac{\sum_{r=b,d} P_{Ui,r}}{1-p_2} \quad (B.14)$$

Simplify the above equation, we have

$$H(U \mid X^{(2)}) = -\sum_{\substack{i=1\\i\in\mathcal{Y}_2}}^{N} P_{Ui}\log\frac{P_{Ui}}{p_2} - \sum_{\substack{i=1\\i\in\mathcal{Y}_3}}^{N} P_{Ui}\log\frac{P_{Ui}}{1-p_2}$$
$$= -\sum_{i=1}^{N} P_{Ui}\log P_{Ui} + p_2\log p_2 + (1-p_2)\log(1-p_2) \quad (B.15)$$
$$= H(U) - H(X^{(2)})$$

Secondly, consider the necessity of the condition.

Using the reduction to absurdity, assuming that for $\exists i \in \mathcal{Y}$, the condition in Eq. (23) are not met, there exists two possible situations according to the basic assumptions:

a) The first situation satisfies: for $\exists j,k \in \mathcal{Y}$,

$$\sum_{r=a,c} P_{Uj,r}\sum_{r=b,d} P_{Uj,r} \neq 0, \sum_{r=a,c} P_{Uk,r}\sum_{r=b,d} P_{Uk,r} \neq 0 \quad (B.16)$$

The extreme point is uncertain, and the domain of $\delta$ is distributed on both sides of the origin:

$$-\min\left(\sum_{r=a,c} P_{Uj,r},\sum_{r=b,d} P_{Uk,r}\right) \leq \delta \leq \min\left(\sum_{r=a,c} P_{Uk,r},\sum_{r=b,d} P_{Uj,r}\right) \quad (B.17)$$

Therefore, if the extreme point is non-zero, then when the extreme point and $\delta$ have different signs, the value of $\Delta H_{jk}^{X^{(2)}}$ is negative; if the extreme point is zero, then when the $\delta$ is non-zero, the value of $\Delta H_{jk}^{X^{(2)}}$ is negative. For various possible extreme points, the value of $\Delta H_{jk}^{X^{(2)}}$ always can be negative, that is, the conditional information entropy of shared data $U$ conditioned on nonprivate characteristic $X^{(2)}$ can be further decreased. Because the information entropy of shared data $U$ being locked to $H(U)$, the mutual information of shared data $U$ and nonprivate characteristic $X^{(2)}$ can be further increased, which contradicts the assumption that the mutual information reaches a maximum value of $H(X^{(2)})$.

b) The second situation satisfies: for $\exists! j \in \mathcal{Y}$,

$$\sum_{r=a,c} P_{Uj,r}\sum_{r=b,d} P_{Uj,r} \neq 0 \quad (B.18)$$

Firstly, it is assumed that for $\exists k \in \mathcal{Y}$, the following condition is satisfied:

$$\sum_{r=a,c} P_{Uk,r} = 0, \sum_{r=b,d} P_{Uk,r} \neq 0 \quad (B.19)$$

The extreme point is

$$\delta^{X^{(2)}*} = \frac{-\sum_{r=b,d} P_{Uk,r}\sum_{r=a,c} P_{Uj,r}}{\sum_{r\in\mathcal{P}} P_{Uj,r} + \sum_{r\in\mathcal{P}} P_{Uk,r}} < 0 \quad (B.20)$$

Therefore, the domain of $\delta$ only remains the non-positive half:

$$-\min\left(\sum_{r=a,c} P_{Uj,r},\sum_{r=b,d} P_{Uk,r}\right) \leq \delta \leq 0 \quad (B.21)$$

If

$$\sum_{r=a,c} P_{Uj,r} \leq \sum_{r=b,d} P_{Uk,r} \quad (B.22)$$

Therefore, the lower bound of $\delta$ is $-P_{Uj,a}-P_{Uj,c}$, and the value of $\Delta H_{jk}^{X^{(2)}}$ when $\delta$ at its lower bound:

$$\Delta H_{jk}^{X^{(2)}}\left(-\sum_{r=a,c} P_{Uj,r}\right) \\ = \sum_{r=b,d} P_{Uj,r}\log\sum_{r=b,d} P_{Uj,r} + \sum_{r=b,d} P_{Uk,r}\log\sum_{r=b,d} P_{Uk,r} \\ -\left(\sum_{r=b,d} P_{Uj,r}+\sum_{r=a,c} P_{Uj,r}\right)\log\left(\sum_{r=b,d} P_{Uj,r}+\sum_{r=a,c} P_{Uj,r}\right) \\ -\left(\sum_{r=b,d} P_{Uk,r}-\sum_{r=a,c} P_{Uj,r}\right)\log\left(\sum_{r=b,d} P_{Uk,r}-\sum_{r=a,c} P_{Uj,r}\right) \quad (B.23)$$

It can be easily proved that for $\forall x,y,z, x+z, y-z \in (0,1)$, the following inequality holds if and only if $x+z \leq y$:

$$x\log x + y\log y \geq (x+z)\log(x+z) + (y-z)\log(y-z) \quad (B.24)$$

Therefore, when Eq. (B.22) holds, the value of $\Delta H_{jk}^{X^{(2)}}$ in the domain of $\delta$ is always non-negative, otherwise, there must be at least one feasible value of $\delta$ that make the value of $\Delta H_{jk}^{X^{(2)}}$ be negative:

$$\sum_{r\in\mathcal{P}} P_{Uj,r} = P_{Uj} \leq P_{Uk} = \sum_{r=b,d} P_{Uk,r} \quad (B.25)$$

If

$$\sum_{r=a,c} P_{Uj,r} \geq \sum_{r=b,d} P_{Uk,r} \quad (B.26)$$

Therefore, the lower bound of $\delta$ is $-P_{Uk,b}-P_{Uk,d}$, and the value of $\Delta H_{jk}^{X^{(2)}}$ when $\delta$ at its lower bound:

$$\Delta H_{jk}^{X^{(2)}}\left(-\sum_{r=b,d} P_{Uk,r}\right) \\ = \sum_{r=a,c} P_{Uj,r}\log\sum_{r=a,c} P_{Uj,r} + \sum_{r=b,d} P_{Uj,r}\log\sum_{r=b,d} P_{Uj,r} \\ -\left(\sum_{r=a,c} P_{Uj,r}-\sum_{r=b,d} P_{Uk,r}\right)\log\left(\sum_{r=a,c} P_{Uj,r}-\sum_{r=b,d} P_{Uk,r}\right) \\ -\left(\sum_{r=b,d} P_{Uj,r}+\sum_{r=b,d} P_{Uk,r}\right)\log\left(\sum_{r=b,d} P_{Uj,r}+\sum_{r=b,d} P_{Uk,r}\right) \quad (B.27)$$

According to Eq. (B.24), when Eq. (B.26) holds, the value of $\Delta H_{jk}^{X^{(2)}}$ in the domain of $\delta$ is always non-negative, otherwise, there must be at least one feasible value of $\delta$ that make the value of $\Delta H_{jk}^{X^{(2)}}$ be negative:

$$\sum_{r=b,d} P_{Uj,r} + \sum_{r=b,d} P_{Uk,r} \leq \sum_{r=a,c} P_{Uj,r} \tag{B.28}$$

Secondly, it is assumed that for $\exists k \in \mathcal{Y}$, the following condition is satisfied:

$$\sum_{r=a,c} P_{Uk,r} \neq 0, \sum_{r=b,d} P_{Uk,r} = 0 \tag{B.29}$$

The extreme point is

$$\delta^{X^{(2)}*} = \frac{\sum_{r=b,d} P_{Uj,r} \sum_{r=a,c} P_{Uk,r}}{\sum_{r\in\mathcal{P}} P_{Uj,r} + \sum_{r\in\mathcal{P}} P_{Uk,r}} > 0 \tag{B.30}$$

Therefore, the domain of $\delta$ only remains the non-negative half:

$$0 \leq \delta \leq \min\left(\sum_{r=a,c} P_{Uk,r}, \sum_{r=b,d} P_{Uj,r}\right) \tag{B.31}$$

If

$$\sum_{r=a,c} P_{Uk,r} \leq \sum_{r=b,d} P_{Uj,r} \tag{B.32}$$

Therefore, the upper bound of $\delta$ is $P_{Uk,a}+P_{Uk,c}$, and the value of $\Delta H_{jk}^{X^{(2)}}$ when $\delta$ at its upper bound:

$$\Delta H_{jk}^{X^{(2)}}\left(\sum_{r=a,c} P_{Uk,r}\right)$$
$$= \sum_{r=a,c} P_{Uj,r} \log \sum_{r=a,c} P_{Uj,r} + \sum_{r=b,d} P_{Uj,r} \log \sum_{r=b,d} P_{Uj,r}$$
$$- \left(\sum_{r=a,c} P_{Uj,r} + \sum_{r=a,c} P_{Uk,r}\right)\log\left(\sum_{r=a,c} P_{Uj,r} + \sum_{r=a,c} P_{Uk,r}\right) \tag{B.33}$$
$$- \left(\sum_{r=b,d} P_{Uj,r} - \sum_{r=a,c} P_{Uk,r}\right)\log\left(\sum_{r=b,d} P_{Uj,r} - \sum_{r=a,c} P_{Uk,r}\right)$$

According to Eq. (B.24), when Eq. (B.32) holds, the value of $\Delta H_{jk}^{X^{(2)}}$ in the domain of $\delta$ is always non-negative, otherwise, there must be at least one feasible value of $\delta$ that make the value of $\Delta H_{jk}^{X^{(2)}}$ be negative:

$$\sum_{r=a,c} P_{Uj,r} + \sum_{r=a,c} P_{Uk,r} \leq \sum_{r=b,d} P_{Uj,r} \tag{B.34}$$

If

$$\sum_{r=a,c} P_{Uk,r} \geq \sum_{r=b,d} P_{Uj,r} \tag{B.35}$$

Therefore, the lower bound of $\delta$ is $P_{Uj,b}+P_{Uj,d}$, and the value of $\Delta H_{jk}^{X^{(2)}}$ when $\delta$ at its lower bound:

$$\Delta H_{jk}^{X^{(2)}}\left(\sum_{r=b,d} P_{Uj,r}\right)$$
$$= \sum_{r=a,c} P_{Uj,r} \log \sum_{r=a,c} P_{Uj,r} + \sum_{r=a,c} P_{Uk,r} \log \sum_{r=a,c} P_{Uk,r}$$
$$- \left(\sum_{r=a,c} P_{Uj,r} + \sum_{r=b,d} P_{Uj,r}\right)\log\left(\sum_{r=a,c} P_{Uj,r} + \sum_{r=b,d} P_{Uj,r}\right) \tag{B.36}$$
$$- \left(\sum_{r=a,c} P_{Uk,r} - \sum_{r=b,d} P_{Uj,r}\right)\log\left(\sum_{r=a,c} P_{Uk,r} - \sum_{r=b,d} P_{Uj,r}\right)$$

According to Eq. (B.24), when Eq. (B.35) holds, the value of $\Delta H_{jk}^{X^{(2)}}$ in the domain of $\delta$ is always non-negative, otherwise, there must be at least one feasible value of $\delta$ that make the value of $\Delta H_{jk}^{X^{(2)}}$ be negative:

$$\sum_{r\in\mathcal{P}} P_{Uj,r} = P_{Uj} \leq P_{Uk} = \sum_{r=a,c} P_{Uk,r} \tag{B.37}$$

In summary, when one of Eqs. (24) and (25) holds, the value of $\Delta H_{jk}^{X^{(2)}}$ is always non-negative, otherwise, there must be at least one feasible value of $\delta$ that make the value of $\Delta H_{jk}^{X^{(2)}}$ be negative.

Although it is not possible to decrease the conditional information entropy of shared data $U$ conditioned on nonprivate characteristic $X^{(2)}$ through direct probability exchange adjustment method, it can be proved that the conditional information entropy of shared data $U$ conditioned on nonprivate characteristic $X^{(2)}$ is greater than $H(U)-H(X^{(2)})$ and has not yet reached the minimum value.

Then, we define set $\mathcal{Y}_6$ as

$$\mathcal{Y}_6 = \left\{i \in \mathcal{Y} \middle| \sum_{r=a,c} P_{Ui,r} \neq 0, \sum_{r=b,d} P_{Ui,r} \neq 0\right\} \tag{B.38}$$

The conditional information entropy of shared data $U$ conditioned on nonprivate characteristic $X^{(2)}$ is

$$H(U \mid X^{(2)})$$
$$= -\sum_{\substack{i=1 \\ i\in\mathcal{Y}_2}}^{N}\left(\sum_{r=a,c} P_{Ui,r}\right)\log\frac{\sum_{r=a,c} P_{Ui,r}}{p_2}$$
$$- \sum_{\substack{i=1 \\ i\in\mathcal{Y}_6}}^{N}\left(\sum_{r=a,c} P_{Ui,r}\right)\log\frac{\sum_{r=a,c} P_{Ui,r}}{p_2} \tag{B.39}$$
$$- \sum_{\substack{i=1 \\ i\in\mathcal{Y}_3}}^{N}\left(\sum_{r=b,d} P_{Ui,r}\right)\log\frac{\sum_{r=b,d} P_{Ui,r}}{1-p_2}$$
$$- \sum_{\substack{i=1 \\ i\in\mathcal{Y}_6}}^{N}\left(\sum_{r=b,d} P_{Ui,r}\right)\log\frac{\sum_{r=b,d} P_{Ui,r}}{1-p_2}$$

Simplify the above equation, we have

$$H(U \mid X^{(2)}) = -\sum_{\substack{i=1 \\ i \in \mathcal{Y}_2}}^{N} P_{Ui} \log \frac{P_{Ui}}{p_2} - \sum_{\substack{i=1 \\ i \in \mathcal{Y}_3}}^{N} P_{Ui} \log \frac{P_{Ui}}{1-p_2}$$

$$- \sum_{r=a,c} P_{Ui,r} \log \frac{\sum_{r=a,c} P_{Ui,r}}{p_2} - \sum_{r=b,d} P_{Ui,r} \log \frac{\sum_{r=b,d} P_{Ui,r}}{1-p_2} \quad \text{(B.40)}$$

$$= -\sum_{\substack{i=1 \\ i \in \mathcal{Y}_2+\mathcal{Y}_3}}^{N} P_{Ui} \log P_{Ui} + p_2 \log p_2 + (1-p_2)\log(1-p_2)$$

$$- \sum_{r=a,c} P_{Ui,r} \log \sum_{r=a,c} P_{Ui,r} - \sum_{r=b,d} P_{Ui,r} \log \sum_{r=b,d} P_{Ui,r}$$

According to Eq. (B.8), we have

$$\sum_{r=a,c} P_{Ui,r} \log \sum_{r=a,c} P_{Ui,r} + \sum_{r=b,d} P_{Ui,r} \log \sum_{r=b,d} P_{Ui,r} < P_{Uj} \log P_{Uj} \quad \text{(B.41)}$$

Therefore, the conditional information entropy of shared data $U$ conditioned on nonprivate characteristic $X^{(2)}$ in Eq. (B.40) satisfies

$$H(U \mid X^{(2)}) > -\sum_{\substack{i=1 \\ i \in \mathcal{Y}_2+\mathcal{Y}_3+\mathcal{Y}_6}}^{N} P_{Ui} \log P_{Ui}$$

$$+ p_2 \log p_2 + (1-p_2)\log(1-p_2) \quad \text{(B.42)}$$

$$= H(U) - H(X^{(2)})$$

Because the information entropy of shared data $U$ being locked to $H(U)$, the conditional information entropy of shared data $U$ conditioned on nonprivate characteristic $X^{(2)}$ is greater than $H(U)-H(X^{(2)})$, which means that the mutual information of shared data $U$ and nonprivate characteristic $X^{(2)}$ is smaller than $H(X^{(2)})$, which contradicts the assumption that the mutual information reaches a maximum value of $H(X^{(2)})$

The two possible situations have been discussed, both of which contradict the assumption that the mutual information reaches a maximum value of $H(X^{(2)})$, then the assumption is incorrect. ∎

### C. Proof of Proposition 3

Considering the symmetry of $j$ and $k$, there are four possible situations:

a) The first situation is $j \in \mathcal{Y}_1, k \in \mathcal{Y}$:

The extreme points of $\Delta H_{jk}^{X^{(1)}}$ and $\Delta H_{jk}^{X^{(2)}}$ are both zero. Because both $\Delta H_{jk}^{X^{(1)}}$ and $\Delta H_{jk}^{X^{(2)}}$ are concave functions that cross the origin and have at most one maximum point, when the extreme points of $\Delta H_{jk}^{X^{(1)}}$ and $\Delta H_{jk}^{X^{(2)}}$ are both zero, the values of $\Delta H_{jk}^{X^{(1)}}$ and $\Delta H_{jk}^{X^{(2)}}$ are always non-positive. Correspondingly, the conditional information entropy of shared data $U$ conditioned on private characteristic $X^{(1)}$ cannot be increased, and the conditional information entropy of shared data $U$ conditioned on nonprivate characteristic $X^{(2)}$ cannot be decreased. Therefore, the mutual information of shared data $U$ and private characteristic $X^{(1)}$ and the mutual information of shared data $U$ and nonprivate characteristic $X^{(2)}$ can only increase simultaneously, that is, the mutual information pair reaches Pareto optimal bound.

b) The second situation is $j, k \in \mathcal{Y}_2 + \mathcal{Y}_3$:

Referring the sufficiency proof of *Propositions 2*, the mutual information of shared data $U$ and nonprivate characteristic $X^{(2)}$ cannot be increased, and the mutual information pair reaches Pareto optimal bound.

c) The third situation is $j, k \in \mathcal{Y}_4$:

Referring the sufficiency proof of *Propositions 1*, the mutual information of shared data $U$ and private characteristic $X^{(1)}$ cannot be decreased, and the mutual information pair reaches Pareto optimal bound.

d) The fourth situation is $j \in \mathcal{Y}_2 + \mathcal{Y}_3, k \in \mathcal{Y}_4$:

Firstly, it is assumed that

$$\sum_{r=a,c} P_{Uj,r} = 0, \sum_{r=b,d} P_{Uj,r} \neq 0 \quad \text{(C.1)}$$

The extreme point is

$$\delta^{X^{(2)}*} = \frac{\sum_{r=b,d} P_{Uj,r} \sum_{r=a,c} P_{Uk,r}}{\sum_{r \in \mathcal{P}} P_{Uj,r} + \sum_{r \in \mathcal{P}} P_{Uk,r}} > 0 \quad \text{(C.2)}$$

Therefore, the domain of $\delta$ only remains the non-negative half:

$$0 \leq \delta \leq \min\left(\sum_{r=a,c} P_{Uk,r}, \sum_{r=b,d} P_{Uj,r}\right) \quad \text{(C.3)}$$

However, the extreme point of $\Delta H_{jk}^{X^{(1)}}$ is uncertain. If the extreme point of $\Delta H_{jk}^{X^{(1)}}$ is zero, the value of $\Delta H_{jk}^{X^{(1)}}$ will be always non-positive, the conditional information entropy of shared data $U$ conditioned on private characteristic $X^{(1)}$ cannot be increased, thus the mutual information of shared data $U$ and private characteristic $X^{(1)}$ cannot be decrease. Therefore, the mutual information pair reaches Pareto optimal bound. If the extreme point of $\Delta H_{jk}^{X^{(1)}}$ is smaller than zero, the value of $\Delta H_{jk}^{X^{(1)}}$ in the domain of $\delta$ will be always non-positive. Similarly, the mutual information pair reaches Pareto optimal bound. If the extreme point of $\Delta H_{jk}^{X^{(1)}}$ is greater than zero, there must be at least one feasible value of $\delta$ that make the value of $\Delta H_{jk}^{X^{(1)}}$ be positive, then we only need to prove that the value of $\Delta H_{jk}^{X^{(2)}}$ in the domain of $\delta$ will be always non-negative. Because the only extreme point of is greater than zero, therefore the minimum value of $\Delta H_{jk}^{X^{(2)}}$ will appear only when $\delta$ is at the edge of its domain. The left-side edge of domain is the origin, and the value of $\Delta H_{jk}^{X^{(2)}}$ when $\delta$ is at the origin is zero, thus we only need to focus on the value of $\Delta H_{jk}^{X^{(2)}}$ when $\delta$ is at the right-side edge of domain.

Secondly, it is assumed that

$$\sum_{r=a,c} P_{Uj,r} \neq 0, \sum_{r=b,d} P_{Uj,r} = 0 \quad \text{(C.4)}$$

The extreme point is

$$\delta^{X^{(2)}*} = \frac{-\sum_{r=b,d} P_{Uk,r} \sum_{r=a,c} P_{Uj,r}}{\sum_{r \in \mathcal{P}} P_{Uj,r} + \sum_{r \in \mathcal{P}} P_{Uk,r}} < 0 \quad \text{(C.5)}$$

Therefore, the domain of $\delta$ only remains the non-positive half:

$$-\min\left(\sum_{r=a,c} P_{Uj,r}, \sum_{r=b,d} P_{Uk,r}\right) \leq \delta \leq 0 \quad \text{(C.6)}$$

Similarly, the extreme point of $\Delta H_{jk}^{X^{(1)}}$ is uncertain. If the extreme point of $\Delta H_{jk}^{X^{(1)}}$ is zero, the value of $\Delta H_{jk}^{X^{(1)}}$ will be always non-positive, the conditional information entropy of shared data $U$ conditioned on private characteristic $X^{(1)}$ cannot be increased, thus the mutual information of shared data $U$ and private characteristic $X^{(1)}$ cannot be decrease. Therefore, the mutual information pair reaches Pareto optimal bound. If the extreme point of $\Delta H_{jk}^{X^{(1)}}$ is greater than zero, the value of $\Delta H_{jk}^{X^{(1)}}$ in the domain of $\delta$ will be always non-positive. Similarly, the mutual information pair reaches Pareto optimal bound. If the extreme point of $\Delta H_{jk}^{X^{(1)}}$ is smaller than zero, there must be at least one feasible value of $\delta$ that make the value of $\Delta H_{jk}^{X^{(1)}}$ be positive, then we only need to prove that the value of $\Delta H_{jk}^{X^{(2)}}$ in the domain of $\delta$ will be always non-negative. Because the only extreme point of is greater than zero, therefore the minimum value of $\Delta H_{jk}^{X^{(2)}}$ will appear only when $\delta$ is at the edge of its domain. The right-side edge of domain is the origin, and the value of $\Delta H_{jk}^{X^{(2)}}$ when $\delta$ is at the origin is zero, thus we only need to focus on the value of $\Delta H_{jk}^{X^{(2)}}$ when $\delta$ is at the left-side edge of domain.

Referring the necessity proof of *Propositions 2*, the value of $\Delta H_{jk}^{X^{(2)}}$ satisfies above conditions when one of Eqs. (24) and (25) holds. ∎